\documentclass[twoside]{IEEEtran}
\usepackage{atbegshi,picture,ifpdf,setspace}
\usepackage[noadjust]{cite}
\usepackage[pdftex]{graphicx}
\usepackage[cmex10]{amsmath}
\usepackage[export]{adjustbox}
\usepackage{xfrac}
\usepackage[caption=false,font=footnotesize]{subfig}
\usepackage{array,tikz,circuitikz,booktabs}
\newlength\imageheight%
\usepackage{dsfont}
\usepackage{stfloats}

\usepackage[english]{babel}
\usepackage[utf8x]{inputenc}
\usepackage[T1]{fontenc}
\usepackage[cmintegrals,varvw]{newtxmath}
\usepackage{acronym}
\usepackage{bm}
\usepackage{siunitx}

\IEEEeqnarraydefcolsep{0}{\leftmargini}
\newcommand*\mat[1]{\textsf{\textit{\textbf#1}}}
\renewcommand*\mat[1]{\bm{#1}}

\makeatletter
\def\ps@IEEEtitlepagestyle{%
  \def\@oddfoot{\mycopyrightnotice}%
  \def\@oddhead{\hbox{}\@IEEEheaderstyle\leftmark\hfil\thepage}\relax
  \def\@evenhead{\@IEEEheaderstyle\thepage\hfil\leftmark\hbox{}}\relax
  \def\@evenfoot{}%
}
\def\mycopyrightnotice{%
  \begin{minipage}{\textwidth}
  \centering \scriptsize
  Copyright~\copyright~2020 IEEE. Personal use of this material is permitted. Permission from IEEE must be obtained for all other uses, in any current or future media, including\\reprinting/republishing this material for advertising or promotional purposes, creating new collective works, for resale or redistribution to servers or lists, or reuse of any copyrighted component of this work in other works by sending a request to pubs-permissions@ieee.org.
  \end{minipage}
}
\makeatother

\AtBeginShipout{\AtBeginShipoutUpperLeft{%
  \put(\dimexpr\paperwidth-1cm\relax,-.55cm){\makebox[0pt][r]{
  \begin{minipage}{\textwidth}
  \centering \scriptsize 
  This is the author's version of an article that has been published in this journal. 
  Changes were made to this version by the publisher prior to publication. 
  \\[0.3ex]
  The final version of record is available at\qquad http://dx.doi.org/10.1109/TAP.2020.3016422
  \end{minipage}
  }}}%
}

\begin{document}
\title{Compact Uniform Circular Quarter-Wavelength Monopole Antenna Arrays with Wideband Decoupling and Matching Networks}
\author{Jonas Kornprobst, \IEEEmembership{Student Member, IEEE}, Thomas J. Mittermaier, \IEEEmembership{Member, IEEE}, Raimund A. M. Mauermayer, \IEEEmembership{Student Member, IEEE}, Gerhard F. Hamberger, Matthias G. Ehrnsperger, \IEEEmembership{Student Member, IEEE}, Bernhard Lehmeyer, Michel T. Ivrla\v c, Ulrik Imberg, Thomas F. Eibert, \IEEEmembership{Senior Member, IEEE}, and Josef A. Nossek, \IEEEmembership{Life Fellow, IEEE}%
\thanks{Manuscript received December 21, 2019; revised June 14, 2020; accepted June 30, 2020; date of this version July 6, 2020. This work was supported in part by the Huawei Technologies Sweden AB, 164 94 Kista, Sweden. \emph{(Corresponding author: Jonas Kornprobst.)}}%
\thanks{J. Kornprobst, T. J. Mittermaier, R. A. M. Mauermayer, G. F. Hamberger,  M. Ehrnsperger and T. F. Eibert are with the Chair of High-Frequency Engineering, Department of Electrical and Computer Engineering, Technical University  of  Munich,  80290  Munich, Germany (e-mail:  j.kornprobst@tum.de, hft@ei.tum.de).}%
\thanks{M. Lehmeyer, M. T. Ivrla\v c and J. A. Nossek are with the Associate Professorship of Signal Processing Methods, Department of Electrical and Computer Engineering, Technical University  of  Munich,  80290  Munich, Germany.}%
\thanks{U. Imberg is with Huawei Technologies Sweden AB, 164 94 Kista, Sweden.}%
\thanks{Color versions of one or more of the figures in this article are availableonline at https://ieeexplore.ieee.org.}%
\thanks{Digital Object Identifier 10.1109/TAP.2020.3016422}
}

\markboth{IEEE Transactions on Antennas and Propagation}%
{Kornprobst \MakeLowercase{\textit{et al.}}: Compact UCAs with Wideband DMNs}

\maketitle

\begin{abstract}
Two novel decoupling and matching networks (DMNs) in microstrip technology for three-element uniform circular arrays (UCAs)  are investigated and compared to a more conventional DMN approach with simple neutralization lines. 
The array elements are coaxially-fed quarter-wavelength monopole antennas over a finite groundplane. 
Three-element arrays are considered since UCAs with an odd number of elements are able to provide an almost constant maximum array factor over the whole azimuthal angular range. 
The new designs are explained from a theoretical point of view and their implementations are compared to four- and three-elements UCAs without DMN in terms of decoupling and matching bandwidth as well as beamforming capabilities. 
In addition to excellent decoupling and matching below $\boldsymbol-\boldsymbol1\boldsymbol6$\,dB, a broader bandwidth is obtained by the two DMNs. 
The reasons for the enhanced bandwidth are similar in both cases: By introducing several circuit elements offering additional degrees of freedom, matching of the monopole input impedances at different frequencies becomes feasible. 
One of the presented designs offers a larger bandwidth, while  the other design is able to provide a better total efficiency. 
Scattering parameters, radiation patterns, beamforming capabilities, and enhanced gain are all verified by measurements over the operating bandwidth.
\end{abstract}


\section{Introduction} %

\IEEEPARstart{F}{uture} wireless communication systems --- such as the next generation mobile wireless network 5G --- strive for ever-increasing data rates for more and more users which are served simultaneously. 
In addition to higher radio frequencies (up to mm-waves) and more bandwidth (e.g., carrier aggregation across frequency bands), one key technology will be to increase the number of antennas in (massive) multiple-input multiple-output (MIMO) systems, and to employ advanced beamforming techniques to serve multiple users with spatial multiplexing~\cite{Han2015,Gupta2015,Muirhead2016,Devoti2016,Heath2016,Busari2018,Roh2014}. 
In particular for the lower GHz range with wavelengths comparable to the size of the mobile devices, it is of utmost importance to keep the size of the mobile antenna array systems reasonably small. 
On the side of the base station, size is not that critical and it might be more beneficial to match and decouple the array elements as good as possible.
In the mm-wave bands, an increasing number of array elements together with
beamforming can be an approach towards a sufficiently large gain \mbox{to overcome the path loss~\cite{Roh2014,Li14,Wei14,Semkin15,Ojaroudiparchin16,Hamberger17,Li2017,Kornprobst2018,Lian2018,Wu2018}.}

In the following, we will focus on compact antenna arrays for the S-band (around $3.6$\,GHz), which is of interest for LTE and 5G and which is already under investigation for various antenna designs~\cite{WRC2015,Spectrum2018,HuaweiSpectrum2018,Al2014,Ban2016,Wong2016,Li2016a,Abdullah2017,WangH2017,Wong2018}. 
Compact arrays are, by their literal meaning, based on very closely spaced antenna elements. 
In particular for a separation distance smaller than half a wavelength, the radiating elements need to be small enough to allow for the narrow spacing. 
Furthermore, super-directive effects might occur, which make efficiency and gain considerations more important than the bare look on the radiation patterns~\cite{Haviland1995,Yaghijan2005,Volmer2009,Sievenpiper2012}. 
Very important is that full beamforming capabilites are still available and no gain loss occurs due to a poor radiation efficiency~\cite{Wang2015b}. 

An important topic of the presented work is to investigate the coupling effects of compact arrays and possibly take advantage of them~\cite{Ivrlac2010,Ivrlac2014,Jeon2017}. 
The coupling effects are tackled by the design and realization of decoupling and matching networks (DMNs). 
Matching commonly means (complex-conjugate)  power matching to a given reference impedance, where 50\,$\Omega$ is a typical widely employed value which is taken throughout this paper. 
Other matching strategies are of course also possible and meaningful, for instance noise matching for low-noise amplifiers (LNAs) in the receive case, where the optimal impedance may even assume complex values~\cite{Lehmeyer2014}.  

For the considered DMN realizations, we have transmit scenarios in mind.
Therefore, matching as well as decoupling levels of $-16$\,dB, i.e., for all scattering parameters $S_{ij}$, 
are requested to reduce the influence of scattered waves onto the transmit power amplifiers (PAs), which may be disturbed in their operating point leading in particular to excitation dependent nonlinearities. 
Moreover, good matching and decoupling is essential for a good efficiency of the system. 
Thereby, decoupling of the array does not mean to cure mutual coupling effects such as modified current distributions on the antenna elements or shadowing effects.
The goal is rather to provide decoupled and matched ports to the transmit power amplifiers. 

Good decoupling is often achieved by orthogonal geometrical configurations~\cite{Yetisir2016,Li2016, Lian2018_2}.
Other possibilities include electromagnetic bandgap and metamaterial structures~\cite{Zhai2016,Ardakani2017}, modified ground structures~\cite{Wang2014,Wang2015,Liu2015,Ban2016,Yu2018}, and coupled parasitic elements~\cite{Soltani2015}.
Also, a lot of literature focuses on two-element (super-directive) arrays~\cite{Andersen1976,Altshuler2005,Arceo2012,Ko2014,Tang2015,Tang2015b}, which are interesting but not sufficient for many applications. 
For more elements, the eigenmode excitations and conjugate-matching techniques are well-known~\cite{Wallace2004,Volmer2008, Wang2013}. 
Tunable DMNs can eliminate the necessity of broadband antenna arrays~\cite{Chen2017}. 

In this paper, we focus on compact uniform circular arrays (UCAs). 
Considering super-directivity, a maximum directivity on the order of $\mathcal O (N^{3/2})$ is possible with UCAs comprising $N$ elements~\cite{Kim2017}.
We investigate three and four element UCAs with quarter-wavelength monopole radiators over a finite ground plane, and demonstrate that three elements are preferable. 
If the maximum beamforming gain in any azimuthal direction is determined for ideal omnidirectional antenna elements, i.e., the array factor, it is found that UCAs with an even number of elements show gain fluctuations while odd-numbered UCAs do not show this behavior and reach a rather constant maximum gain in all azimuthal directions. 
Interesting to note is that the directivity into the minimum-gain directions of an even-numbered UCA can be realized in all directions using an odd-numbered UCA with just one element less. 
Related results are found in literature, but the behavior has not been studied in detail~\cite{Durrani2004,Belloni2006,Fallahi2007}. 

In our study, the employed monopole antenna elements are fed from the rear side of the ground plane via a coaxial transmission line, offering the opportunity to implement decoupling and matching networks in microstrip technology. 

The paper is organized as follows. In Section II, the performance differences of UCAs with even and odd numbers of elements are investigated with focus on the three and four element cases. 
Then, three different DMNs are analyzed in Section~III from a theoretical point of view for the three-element scenario: 
The first one is a state-of-the-art approach with rather simple neutralization transmission lines. 
The second one is a rather complicated two-stage network with the largest decoupling and matching bandwidth, but with the downside of a slightly reduced radiation efficiency. 
The second one features several decoupling paths in a star-triangle configuration and offers a superior efficiency with a slightly increased bandwidth.
The specific design procedure for the simulation, the fabrication, and scattering parameter measurements for these three DMNs is discussed in Section~IV.
Finally, all simulations of the DMN designs are verified by measurements in Section~V, for radiation patterns, gain, and beamforming vectors.

\section{UCAs with Even and Odd Numbers of Elements} %

Towards the design of a compact UCA with an optimal number of elements, we investigate some 
general properties of ideal UCAs  first. 
In the following array factor considerations, it becomes clear why a UCA with an even number of elements should be replaced by a UCA with one element less. 
After introducing an antenna setup based on monopoles, which is used throughout this work, the claims are also validated by full-wave simulations and measurements. 

\subsection{Theoretical Investigations with Ideal Radiators}

\begin{figure}[tp]
\centering
\includegraphics[]{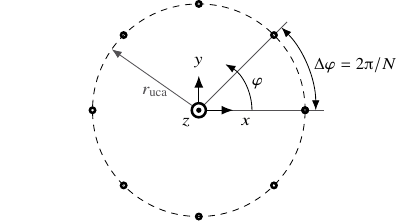}
\caption{Sketch of a UCA in the \textit{xy}-plane.\label{fig:firstsetup}}
\end{figure}

We consider a scenario with ideal radiators for an array-factor description, where we assume a suppressed $\mathrm e ^{\,\mathrm j \omega t}$ time dependence with angular frequency $\omega$ throughout this paper. 
An ideal $N$-element UCA on a circle with radius $k_0 r_\mathrm{uca}=0.2\uppi$ is considered, see Fig.~\ref{fig:firstsetup}, 
and we obtain the array factor~\cite{Balanis2005}
\begin{equation}
\mathit{AF} = \sum\nolimits_{n=0}^{N-1} a_n \mathrm e^{\,\mathrm j k_0r_\mathrm{uca}
\sin\vartheta(\cos \varphi \cos({2 \uppi}n/{N})+ \sin \varphi   \sin({2 \uppi}n/{N}))}
\end{equation}%
with the complex element excitations $a_n$, the free-space wavenumber $k_0=\omega\sqrt{\varepsilon_0\mu_0}=2\uppi/\lambda_0$, and the corresponding free-space wavelength $\lambda_0$.
The array elements are assumed to be quarter-wavelength monopole antennas oriented along the $z$-axis. 
Their radiation characteristic\,---\,idealized without mutual influence\,---\,reads~\cite{Balanis2005}
\begin{equation}
C(\vartheta,\varphi)= 
\begin{cases} 
\cos(\uppi/2\cos\vartheta)/\sin\vartheta & 0\le\vartheta<\uppi/2\\
0             & \uppi/2\le\vartheta\le\uppi\\
\end{cases}\,.
\end{equation}

\begin{figure}[!t]
\centering
\includegraphics[]{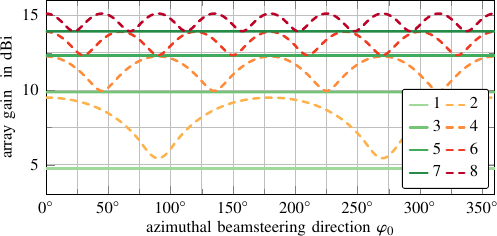}
\caption{Array gain of UCAs with different numbers of idealized quarter-wavelength monopoles with optimal azimuthal beamsteering in the $\vartheta=70^\circ$ cut.\label{fig:ag1}}
\end{figure}

With the knowledge of this setup, we can calculate optimal beamsteering vectors for each direction in space according to~\cite{Ivrlac2010a}.
Considering the $\vartheta=\SI{70}{\degree}$ cut, we determine the maximum array gain for each azimuthal direction~$\varphi_0$  as the product of single-element gain and array factor. 
These gain graphs are shown in Fig.~\ref{fig:ag1}
for UCAs with 1 to 8 elements and radius $r_\mathrm{uca}=0.1\lambda_0$.
We observe, as announced in the introduction that the achievable array gains\footnote{Speaking of gain, we refer to the realized gain throughout this paper both for simulation and measurement results. For the theoretical investigations with uncoupled radiators, the realized gain equals the IEEE gain and the directivity.} with an odd number $N$ of elements are independent of the desired steering angle. However, for even numbers $N$, the achievable gain fluctuates between the values achievable with $N-1$ and $N+1$ elements. 

So far, gain is equal to directivity. 
Major differences will be observed not only with non-ideal radiation efficiencies and with non-ideal non-omnidirectional patterns, but also if the input impedances of the array antennas are considered together with the active reflection coefficients dependent on the different beamforming vectors. 
Direction-dependent gain drops may be observed due to the varying mismatch. 

\begin{figure}[tp]
\centering
\begin{minipage}[t]{0.475\linewidth}
 \subfloat[\hspace*{1.5cm}]{\includegraphics{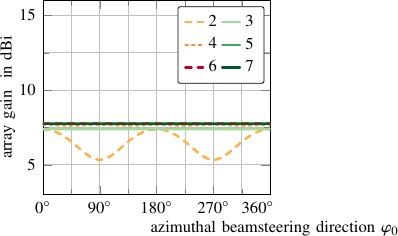}}%
\end{minipage}%
\hspace*{\fill}%
\begin{minipage}[t]{0.475\linewidth}
 \raggedleft
 \subfloat[]{\includegraphics{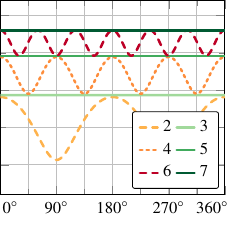}}%
\end{minipage}%
\caption{Array gain of UCAs with different numbers of minimum-scattering quarter-wavelength monopoles. (a)~No matching circuit. (b)~Ideal DMN for power matching.\label{fig:gain_PM}}
\end{figure}

The last non-ideality is integrated into the theoretical analysis, bringing the array model one step closer to reality. 
The individual monopole antennas are now canonical minimum scattering antennas. 
The azimuthal gain pattern is still omnidirectional, even though coupling impedances between the antenna elements are observed~\cite{Ivrl2017}.
Next, we implement power matching with a DMN, which is assumed to be perfect for now. 
The gain in the $\vartheta=70^\circ$ cut with and without DMN is shown in Fig.~\ref{fig:gain_PM}.
Other $\vartheta$-cuts show the same behavior, but are influenced by the varying single-radiator gain.
The gain (or better directivity) curves from Fig.~\ref{fig:ag1} are exactly reproduced with the DMN. Without DMN, the mismatch loss reduces the gain to small values, where the different numbers of UCA-elements are almost indistinguishable.
The model employed for Fig.~\ref{fig:gain_PM} does not consider reduced radiation efficiencies and distorted radiation patterns due to closely placed radiators.

\subsection{Setup of Real-World Implementations}

In order to justify the three-element approach for the subsequent designs also by full-wave simulations and measurements, we describe the design, fabrication, and measurement of three-element and four-element compact antenna arrays without decoupling network.
The described monopole configuration exhibits horizontally omnidirectional beamforming capabilities, but a fixed directivity and beamwidth in the vertical patterns. 

\begin{figure}[tp]
 \centering%
\begin{minipage}[t]{0.475\linewidth}
  \subfloat[]{\includegraphics{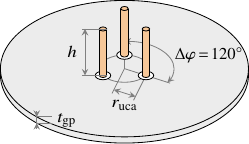}}
\end{minipage}
\hspace*{0.32cm}%
\begin{minipage}[t]{0.475\linewidth}
  \subfloat[]{\includegraphics{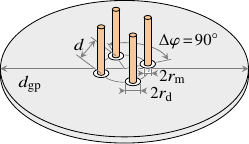}}
\end{minipage}
\\%
\subfloat[]{\includegraphics{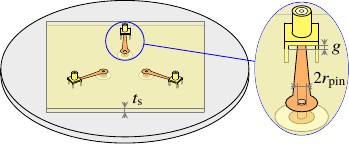}}%
\hspace*{\fill}%
\subfloat[]{\includegraphics{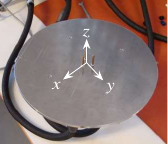}}
\caption{Sketch of UCAs built with monopole antennas over a finite circular groundplane. (a)~Three elements. (b)~Four elements. (c)~Back side with feeding, no DMN. (d)~Photograph of built three-element array.\label{fig:UCAs}}
\end{figure}

The UCAs consisting of quarter-wavelength monopoles are located on a circle of a certain radius~$r_\mathrm{uca}$, see Fig.~\ref{fig:UCAs}. 
The distance between the array elements
\begin{equation}
d = r_\mathrm{uca}\sqrt{2-2\cos \upDelta\varphi}
\end{equation} 
is known from geometrical considerations, where $\upDelta\varphi$ is the angle between the antennas. 
The monopoles on the circular metallic ground plane with diameter $d_\mathrm{gp}=100\,\textrm{mm}$ and thickness of $t_\mathrm{gp}=5\,\textrm{mm}$ exhibit a  radius $r_\mathrm m=1.5\,\textrm{mm}$ and  a height $h\approx0.25\lambda_0$, with the free-space wavelength $\lambda_0$, where $h$ is carefully adjusted in each realization for a resonance frequency of $3.6\,$GHz. 
The coordinate system is defined by the positive $z$-axis pointing into the direction of the monopoles, away from the ground plane. 
The feeding through the groundplane is realized in a coaxial fashion: 
the outer coaxial radius is $r_\mathrm{d}=2.5\,\textrm{mm}$ and the dielectric is Teflon with relative permittivity $\varepsilon_\mathrm r=2.2$. 
This results in a characteristic line impedance of about\ 20$\,\Omega$, which matches the input impedance of the monopoles and does not reduce the impedance bandwidth. 
The theoretically expected monopole input impedance of $36.5$\,$\Omega$~\cite{Balanis2005} is reduced because of the monopole thickness and the array coupling. 
On the rear side of the groundplane, the feed transitions into a microstrip structure on a low-loss RO3003 substrate~\cite{RO3003} of thickness $t_\mathrm s$ and $\varepsilon_\mathrm r=3$ via a pin of radius $r_\mathrm{pin}=0.5\,\textrm{mm}$, which is soldered to a copper matching disk. 
The capacitance of this disk can compensate for the inductive behavior of the coaxial-microstrip transition and is tuned for each realization of the UCA. 
Except otherwise mentioned, $t_\mathrm s = 0.508\,$mm. 
For the following UCAs without DMN, a linear taper matches the impedance to 50$\,\Omega$. 
Measurements are conducted at an SMA connector, which is placed with a defined distance of $g=\SI{1}{mm}$ above the microstrip line. 

\subsection{Characterization of the Three-Element Prototype}

\begin{figure}[tp]
\centering
\includegraphics[]{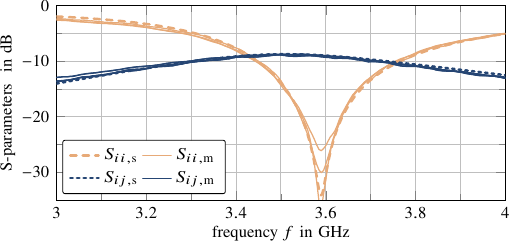}
\caption{Scattering parameters of three-element monopole array.\label{fig:sparam_3}}
\end{figure}

The three-element monopole array was built with a distance $d=0.17\lambda_0$ (i.e., $r_\mathrm{uca}=0.1\lambda_0$) and $h=0.22\lambda_0$. 
A photograph of this prototype during S-parameter measurement is shown in Fig.~\ref{fig:UCAs}(d). 
The S-parameters of this array are given in Fig.~\ref{fig:sparam_3}, where full-wave simulation results and measurement results from a vector network analyzer (VNA) show excellent agreement.  
All simulations were performed with the frequency domain solver of Computer Simulation Technology Microwave Studio (CST MWS)~\cite{CST}.
The input matching of all three ports reaches well below $-20$\,dB at $3.6$\,GHz, and the $-16$-dB bandwidth is larger than 150\,MHz. 
The coupling between the array feeding ports reaches up to about $-9$\,dB; a value considered too large for practically used antenna arrays. 

\begin{figure}[tp]
\centering%
\begin{minipage}[t]{0.475\linewidth}
 \subfloat[]{\includegraphics[]{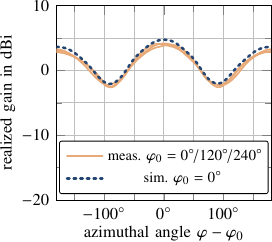}}%
\end{minipage}
\hspace*{\fill}%
\begin{minipage}[t]{0.475\linewidth}
 \raggedleft
 \subfloat[]{\includegraphics[]{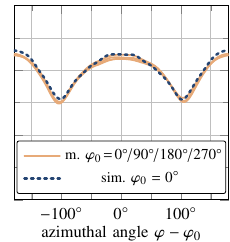}}%
\end{minipage}
\\%
\subfloat[]{\includegraphics[]{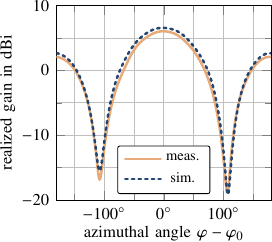}}%
\hspace*{\fill}%
\subfloat[]{\includegraphics[]{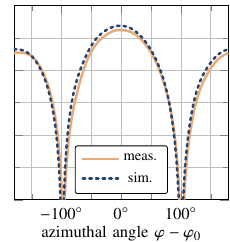}}%
\caption{Comparison of simulated and measured active element and array patterns at a $\vartheta=70^\circ$ cut. (a)~Meas. three-element UCA individual patterns. (b)~Meas. four-element UCA individual patterns. (c)~Three-element UCA for beamsteering in $\varphi_0= 300^\circ$ direction. (d) Four-element UCA for beamsteering in $\varphi_0= 300^\circ$ direction.\label{fig:t1-11abc}}
\end{figure}

\begin{figure}[tpb]
\centering
\begin{minipage}[t]{0.5\linewidth}
 \subfloat[]{\includegraphics[height=3.25cm]{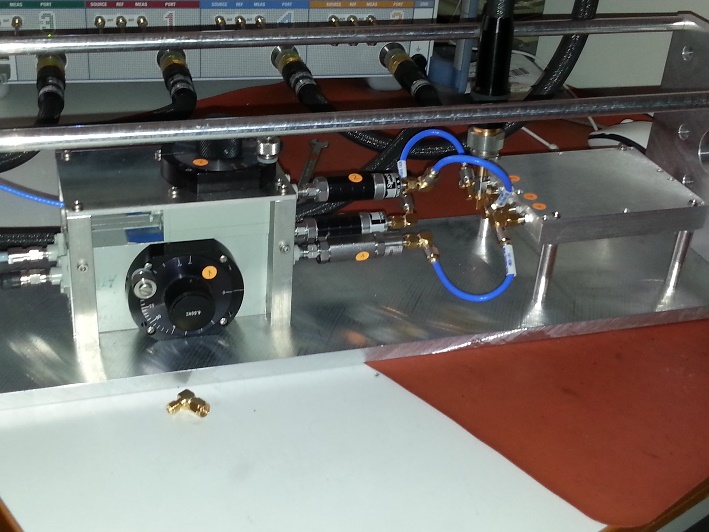}}%
\end{minipage}
\hfill%
\begin{minipage}[t]{0.45\linewidth}
 \raggedleft
 \subfloat[]{\includegraphics[height=3.25cm]{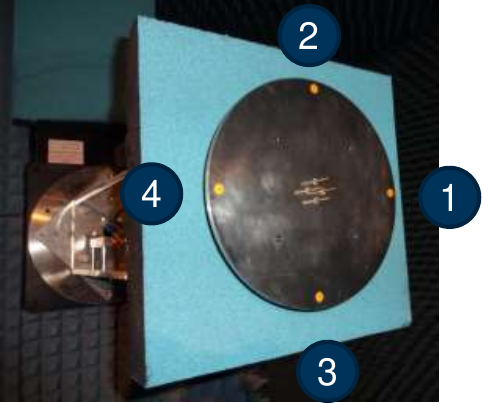}}%
\end{minipage}
\caption{UCA measurement setup. (a)~Analog beamforming network during VNA measurement. (b)~Four-element UCA mounted in anechoic chamber.\label{fig:t1-8abc}}
\end{figure}

The radiation and total efficiencies are $-0.01$\,dB and $-1.28$\,dB in the simulation. 
The poor total efficiency is explained by the mutual array element interactions. 
The simulated single element patterns show a realized gain of $5.09$\,dBi. 
The realized gain is enhanced to values between $6.5$\,dBi and $7$\,dBi if optimal beamforming based on radiation patterns and S-parameters is performed, i.e., the azimuthal gain is maximized in the $\vartheta=70^\circ$-cut. 
The far fields were measured in the anechoic chamber of the Chair of High-Frequency Engineering, Technical University of Munich and the resulting gain radiation patterns are shown in Fig.~\ref{fig:t1-11abc}~\cite{Eibert2005,Neitz2017}. %
Optimal beamforming weights for the $\vartheta=70^\circ$ cut were calculated for an exemplary azimuthal beamsteering angle of $\varphi_0=300^\circ$ and fed into the array via an analog beamforming network, see Fig.~\ref{fig:t1-8abc}. 
The measurement results are corrected for the measured efficiency of the feeding network to obtain the realized gain of the array with the beamforming vector. 
The attained patterns agree very well with the corresponding patterns from CST MWS simulation, see Fig.~\ref{fig:t1-11abc}(c); 
the gain deviation is explained by the measurement uncertainties and also by deviations of the measurement setup as compared to the simulation model, e.g., the imperfect modeling of the absorbers behind the ground plane as displayed in Fig.~\ref{fig:t1-8abc}(b).

\subsection{Characterization of the Four-Element Prototype}

The four-element UCA was designed to have the monopoles placed on a circle with the same radius $r_\mathrm{uca}=0.1\lambda_0$ as for the one with three elements. 
Therefore, the element distance is now slightly smaller with $d=0.14\lambda_0$ and $h$ is still $0.22\lambda_0$.

\begin{figure}[tp]
\centering
\includegraphics[]{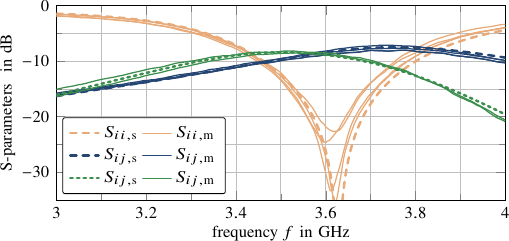}
\caption{Scattering parameters of four-element monopole array. The first, blue $S_{ij}$ crosstalk curve is for neighbouring-element crosstalk, e.g., $S_{12}$, while the second, green one is for opposite array elements, e.g., $S_{14}$, see Fig.~\ref{fig:t1-8abc}(b).\label{fig:t1-14a}}
\end{figure}

A photograph of the fabricated prototype is shown in Fig.~\ref{fig:t1-8abc}(b), where the prototype is mounted on the positioner in the antenna measurement chamber. 
The scattering parameters of the prototype are seen in Fig.~\ref{fig:t1-14a}, where full-wave simulation results and measurement results are compared with good agreement. 
The input matching of the four ports is again very good, but the coupling $S_{ij}$ between the ports is a bit stronger (both in simulation and measurement) than in the case of the three-element array. 
Also, three coupling curves are observed for each fed port: two identical ones for neighboring elements and one for elements across the diagonal.

The radiation and total efficiencies are $-0.06$\,dB and $-2.46$\,dB in simulation, which is worse than for the three-element array.
Hence, the simulated  realized gain drops to $3.84$\,dBi. 
The measured gain patterns are given in Fig.~\ref{fig:t1-11abc}(b).
In Fig.~\ref{fig:t1-11abc}(d), the beamformed gain pattern for the exemplary direction $\varphi_0=300^\circ$ goes up to a very similar level as for the three-element UCA.

\subsection{Beamforming Gain for Azimuthal Scans with Three- and Four-Element UCAs}

\begin{figure}[tp]
\centering
\includegraphics[]{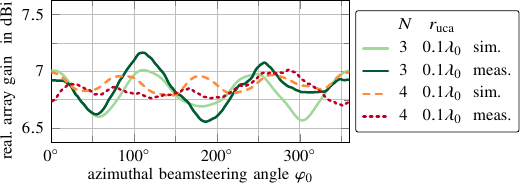}
\caption{Optimum array gain dependent on the scan angle for simulations and measurements of  UCAs with $N=3$ and $4$ elements and with $r_\mathrm{uca}=0.1\lambda_0$.\label{fig:gainreal}}
\end{figure}

To summarize the investigations of Section II, the azimuth angle dependent beamforming behavior for a fixed polar angle cut of $\vartheta=70^\circ$ is investigated for all so-far presented UCAs in Fig.~\ref{fig:gainreal}. 
The comparison involves simulations and measurements of the three- and four-element UCAs with radius of $0.1\lambda_0$. 
For the angle $300^{\circ}$\!\!, the displayed gains agree with those shown in Figs.~\ref{fig:t1-11abc}(c) and (d).

Dependent on the radiation angle, the optimized array gains exhibit some minor oscillations even for the three-element UCA, which is due to the variations in the mismatch and in the single-element patterns of the array (influenced by the finite groundplane, the backlobes, and the mainlobe direction).
This effect should be considered separately from the array factor oscillations studied in this section. 
The array gain of all UCAs is on the same level just below 7\,dBi, which is  a bit lower than observed in Fig.~\ref{fig:gain_PM}. 
It remains to study how real DMNs influence the gain, in particular towards the achivable bandwidth and efficiency.

As outlined in  Section IIA, UCAs with odd numbers of array elements should be preferred over even numbered ones, if each possible beamsteering direction is equally important for the realizable array gain. 

\section{Theoretical Analysis of Decoupling and Matching Networks\label{sc3}} %

\begin{figure}[tp]
\centering
\vspace*{-0.2cm} %
\subfloat[]{\includegraphics{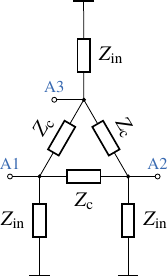}}%
\hfill
\subfloat[]{\includegraphics{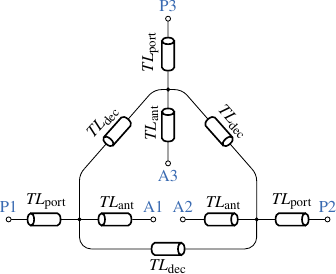}}
\caption{Schematic of the neutralization lines DMN. For the sake of readability, the ground connections of all ports are omitted in this Figure. (a)~Equivalent circuit of the symmetric three-element UCA; this circuit matches an impedance-matrix description of the UCA. (b)~Neutralization lines DMN. The equivalent circuit of (a) is connected at the ports A1--A3.\label{fig:declin}}
\end{figure}

In the following, three DMNs for the presented three-monopole UCA are discussed. 
Since the environment of each monopole is the same, the three antenna excitation ports A1--A3 form a symmetric three-port, see Fig.~\ref{fig:declin}(a), which is for instance modeled with input impedances $Z_\mathrm{in}$ and coupling impedances $Z_\mathrm{c}$. 
We describe three designs of a decoupling and matching network (DMN) that can be used in conjunction with such a symmetric three-port; two of them show the very same three-port symmetry and the third exhibits a rotational symmetry. 
In principle, all of them match $Z_\mathrm{in}$ to $Z_0=50\,\Omega$ and  minimize the influence of $Z_\mathrm c$. 
For the theoretical analysis, fixed input and coupling impedances are assumed.

\subsection{DMN with Neutralization Transmission Lines}

Realizing a DMN with neutralization lines is a well-known technique~\cite{Wang2013}; we briefly summarize the concept in the following and explain the special considerations of our implementation. 
In Fig.~\ref{fig:declin}(b), the basic structure of this kind of DMN is shown. 
First, the antenna modelled as in Fig.~\ref{fig:declin}(a) is connected to the ports A1--A3. 
The coupling is eliminated by the transmission lines $\mathit{TL}_\mathrm{dec}$ connecting the antenna ports. 
Additionally, the lines $\mathit{TL}_\mathrm{ant}$ allow for impedance and phase adjustments. 
Second, the connection and matching to the \mbox{50-$\Omega$} ports P1--P3 is realized by the transmission line $\mathit{TL}_\mathrm{port}$. 
Exploiting the symmetry of the UCA, the DMN is also symmetric. 

\subsection{Two-Stage DMN}

\begin{figure}[tp]
\centering
\includegraphics{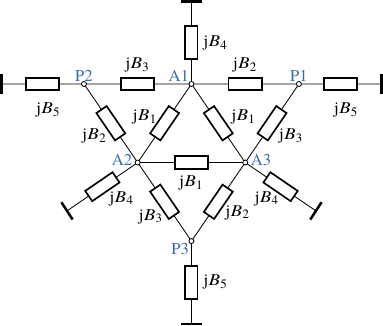}
\caption{Network structure of the proposed decoupling and matching six-port. All ports have ground connections which are not shown. A1--A3 are the antenna ports and P1--P3 are the decoupled and matched feeding ports.}
\label{fig_proposed}
\end{figure}

For the next design, we start with a lumped-element decoupling circuit which shows a rotational symmetry. 
We will demonstrate the ability of this network to decouple and match a symmetric three-port and derive an equivalent circuit with transmission lines for a single frequency. 
The proposed DMN circuit  is a linear, reciprocal, and (ideally) loss-less DMN six-port, see Fig.~\ref{fig_proposed}. 
If the ports A1--A3 are connected to the antenna array excitation ports, the ports P1--P3 are decoupled and present a prescribed impedance, i.e., they are matched. 
The design of the DMN can be realized using planar micro-strip technology: only connections  which do not lead to cross-overs of transmission lines are involved.
The DMN has already been presented in~\cite{lehmeyer2018receiver}.

The proposed DMN circuit exhibits five real-valued degrees of freedom\,---\,the susceptances $B_1$ to $B_5$. 
In total, it consists of 15 loss-less reactances.
Numbering the ports A1--A3 and P1--P3 from 1 to 6, the DMN is described by the admittance matrix
\begin{equation}
\mat{Y}_\mathrm{DMN}=\mathrm{j}\begin{bmatrix}\mat{A}&\mat{B}^\mathrm{T}\\\mat{B}&\mat{C}\end{bmatrix}\,,
\label{e_eins}
\end{equation}
with the real-valued sub-matrices
\begin{equation}
\mat{A}=\begin{bmatrix}2B_1+\sum\nolimits_{k=2}^4 B_k&-B_1&-B_1\\-B_1&2B_1+\sum\nolimits_{k=2}^4 B_k&-B_1\\-B_1&-B_1&2B_1+\sum\nolimits_{k=2}^4 B_k\end{bmatrix}\,,
\label{e_zwei}
\end{equation}
\begin{equation}
\mat{B}=\begin{bmatrix}-B_2&0&-B_3\\-B_3&-B_2&0\\0&-B_3&-B_2\end{bmatrix}\,,
\label{e_drei}
\end{equation}
\begin{equation}
\mat{C}=(B_2+B_3+B_5)\mat 1_3\,,
\label{e_vier}
\end{equation}
where $\mat 1_3$ is the $3\times3$ identity matrix.

Connecting the antenna array at A1--A3 (modeled by the admittance matrix $\mat{Y}_{\!\mathrm A}$, see Fig.~\ref{fig:declin}(a)), we obtain the admittance matrix
\begin{equation}
\mat{Y}=\mathrm{j}\mat{C}+\mat{B}\,(\mat{Y}_{\!\mathrm A}+\mathrm{j}\mat{A})^{-1}\mat{B}^\mathrm{T}\,.
\label{e_funf}
\end{equation}
Since the antenna array is symmetric, $\mat{Y}_{\!\mathrm A}$ has the structure
\begin{equation}
\mat{Y}_\mathrm{A}=\mat{Z}_\mathrm{A}^{-1}=\left[\begin{array}{ccc}\alpha&\beta&\beta\\\beta&\alpha&\beta\\\beta&\beta&\alpha\end{array}\right]\,,\qquad\textrm{with } \alpha,~\beta~\in \mathds{C} \,\Omega^{-1}\,,
\label{e_sechs}
\end{equation}
which depends on the symmetric antenna array input and coupling impedances as 
\iffalse
\begin{equation}
\mat{Y}_\mathrm{A}=\frac{1}{Z_\mathrm{in}^2+Z_\mathrm{in}Z_\mathrm{c}-2Z_\mathrm{c}^2}
\left[
\begin{array}{ccc}
\!\!Z_\mathrm{in}\!+\!Z_\mathrm c\!\!&-Z_\mathrm{c}&-Z_\mathrm{c}\\
-Z_\mathrm{c}&\!\!Z_\mathrm{in}\!+\!Z_\mathrm c\!\!&-Z_\mathrm{c}\\
-Z_\mathrm{c}&-Z_\mathrm{c}&\!\!Z_\mathrm{in}\!+\!Z_\mathrm c\!\!
\end{array}\right].
\label{e_sechsa}
\end{equation}
\else
\begin{equation}
\mat{Y}_\mathrm{A}=\frac{1}{Z_\mathrm{in}Z_\mathrm{c}}
\left[
\begin{array}{ccc}
\!\!2Z_\mathrm{in}\!+\!Z_\mathrm c\!\!&-Z_\mathrm{in}&-Z_\mathrm{in}\\
-Z_\mathrm{in}&\!\!2Z_\mathrm{in}\!+\!Z_\mathrm {in}\!\!&-Z_\mathrm{in}\\
-Z_\mathrm{in}&-Z_\mathrm{in}&\!\!2Z_\mathrm{in}\!+\!Z_\mathrm c\!\!
\end{array}\right].
\label{e_sechsa}
\end{equation}
\fi
Choosing the matrix $\mat A$ 
\begin{equation}
\mat A = -\mathrm{Im}\{\mat Y_\mathrm{\! \mathrm A}\}\label{eq:imagY}
\end{equation}
as the negative imaginary part of $\mat{Y}_{\!\mathrm A}$
and using~\eqref{e_zwei} and \eqref{e_sechs}, 
we obtain the conditions
\begin{equation}
B_1=\mathrm{Im}\{\beta\} \,,
\label{e_achta}
\end{equation}
\begin{equation}
B_2+B_3+B_4=-\mathrm{Im}\{\alpha+2\beta\}\,.
\label{e_achtb}
\end{equation}
Furthermore, \eqref{e_funf} reduces to
\begin{equation}
\mat{Y}=\mathrm{j}\mat{C}+\mat{B}\,(\mathrm{Re}\{\mat{Y}_{\!\mathrm A}\}^{-1})\mat{B}^\mathrm{T}\,.
\label{e_neun}
\end{equation}
Expressing the inverse of the real part of $\mat Y_\mathrm{A}$ leads to
\begin{equation}
\left(\mathrm{Re}\{\mat{Y}_{\!\mathrm{A}}\}\right)^{-1}=\begin{bmatrix}a&b&b\\b&a&b\\b&b&a\end{bmatrix}\,,\quad\textrm{with } a,b\in\mathds{R}\,\Omega\,.
\label{e_zehn}
\end{equation}
It follows that
\begin{equation}
\mat{B}\left(\mathrm{Re}\{\mat{Y}_{\!\mathrm{A}}\}\right)^{-1}\mat{B}^\mathrm{T}=\begin{bmatrix}\gamma&\xi&\xi\\\xi&\gamma&\xi\\\xi&\xi&\gamma\end{bmatrix}
\label{e_elf}
\end{equation}
with the two entries $\gamma$ and $\xi$ with
\begin{equation}
\xi=B_2B_3a+\left(B_2^2+B_2B_3+B^2_3\right)b\,.
\label{e_zwolf}
\end{equation}
For port decoupling, $\xi=0$ must hold and gives a condition for $B_3$ in the form of
\begin{equation}
B_3=-B_2\frac{a+b\pm\sqrt{a^2+2ab-3b^2}}{2b}\,.
\label{e_vierzehn}
\end{equation}
Since $B_3$ must be real-valued, the expression under the square-root must be positive, i.e.,
\begin{equation}
a^2+2ab-3b^2\geq0\,.
\label{e_funfzehn}
\end{equation}
This is fulfilled for compact arrays: In our investigations, this was the case for $d \gtrsim 0.2\lambda$.
With~\eqref{e_vierzehn} and \eqref{e_elf}, we get 
\begin{equation}
\gamma=B_2^2\frac{\left(a^2+ab-2b^2\right)\left(a+b\pm\sqrt{a^2+2ab-3b^2}\right)}{2b^2}\,.
\label{e_sechzehn}
\end{equation}
In order to obtain a matched circuit, the desired form of~\eqref{e_neun} looks like the diagonal matrix\footnote{In this work, we only consider matching to a real impedance. 
Nevertheless, the DMN is easily adapted by changing~\eqref{e_siebzehn} for any complex PA output or LNA input impedance if desired.}%
\begin{equation}
\mat{Y}=Z_0^{-1}\mat 1_3\,,
\label{e_siebzehn}
\end{equation}
where $Z_0$ is the prescribed value of the impedance of the decoupled ports. 
This gives the two further conditions
\begin{equation}
\gamma=Z_0^{-1}\,,
\label{e_achtzehn}
\end{equation}
\begin{equation}
B_5=-B_2-B_3\,.
\label{e_zwanzig}
\end{equation}
Eq.~\eqref{e_sechzehn} is rewritten as
\begin{equation}
B_2=\pm\sqrt{\frac{2b^2 Z_0^{-1}}{\left(a^2+ab-2b^2\right)\left(a+b\pm\sqrt{a^2+2ab-3b^2}\right)}}\,\,,
\label{e_neunzehn}
\end{equation}
where the $+$ or $-$ signs in the  two $\pm$ symbols can be chosen independently. Hence, there are up to
four solutions for $B_2$, and we choose whichever is more convenient with respect to the practical realization of the circuit. 
Having $B_2$, $B_3$ is obtained from \eqref{e_vierzehn}. $B_1$ is already known from~\eqref{e_achta}, and the final two values are given by~\eqref{e_achtb} and~\eqref{e_zwanzig}.

\begin{figure}[tp]
\centering
\subfloat[]{\includegraphics{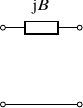}}%
\hspace*{.9cm}
\subfloat[]{\includegraphics{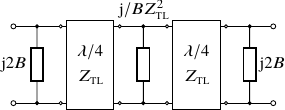}}
\caption{Modeling reactive lumped elements with transmission lines. (a)~Series reactance. (b)~Equivalent model with $\lambda/4$ transmission lines.\label{fig:ts_eq}}
\end{figure}

\begin{figure}[tp]
\centering
\includegraphics{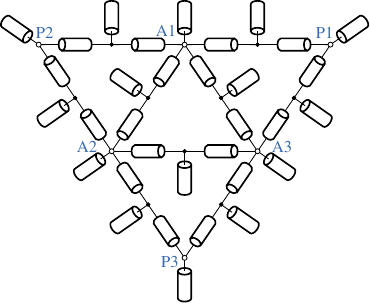}
\caption{Two-stage DMN from Fig.~\ref{fig_proposed} realized with (possibly planar) transmission lines without intersections. Ground connections are omitted for readability. Stubs represent susceptances  at the design frequency and serial lines are quarter-wavelength transformers.\label{fig:DMN_ts_tl}}
\end{figure}

The next step is to find an equivalent transmission line circuit for the proposed lumped-element DMN. 
Therefore, we replace the floating lumped elements as depicted in Fig.~\ref{fig:ts_eq}(a) by a 
the circuit in Fig.~\ref{fig:ts_eq}(b).
We find that both are equivalent at the single frequency where the length of the transmission lines is equal to $\lambda/4$. 
The remaining task  is to replace all shunt reactances by transmission lines, which is easily done by open stubs with the well-known input admittance
\begin{equation}
Y_\textsc{tl} = \mathrm j Z_\textsc{tl}^{-1}\tan(k_\textsc{tl} l_\textsc{tl})\,,
\end{equation}
with the wavenumber $k_\textsc{tl}=2 \uppi / \lambda$ and the physical line length $l_\textsc{tl}$.
The final step to calculate the electrical line length and impedances, and subsequently microstrip dimensions, was carried out for the inital designs before further optimization.
Without going into detail about all transmission line values, the final DMN is given in Fig.~\ref{fig:DMN_ts_tl}.
It features 33 pieces of transmission lines, each with a length of about a quarter wavelength.

\subsection{Star-Triangle DMN}

\begin{figure}[tp]
\centering
\includegraphics{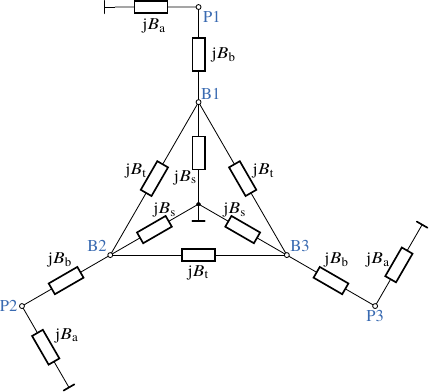}
\caption{Star-triangle DMN, port ground omitted for readability.\label{fig:DMN_lehma}}
\end{figure}

We follow the same procedure as in Section~\ref{sc3}\emph{B}, starting with a symmetric, lossless and reciprocal sixport consisting of 12 ideal susceptances, see Fig.~\ref{fig:DMN_lehma}. The DMN has already been presented in~\cite{lehmeyer2018receiver}.%

We have again for the admittance matrix the same expression given in~\eqref{e_eins} and~\eqref{e_neun} but now with
\begin{equation}
\mat{A}=\begin{bmatrix}
 2B_\mathrm t+B_\mathrm s+B_\mathrm b&-B_\mathrm t&-B_\mathrm t\\
 -B_\mathrm t&2B_\mathrm t+B_\mathrm s+B_\mathrm b&-B_\mathrm t\\
 -B_\mathrm t&-B_\mathrm t&2B_\mathrm t+B_\mathrm s+B_\mathrm b
\end{bmatrix}\,,
\end{equation}
\begin{equation}
\mat{B}=-B_\mathrm b \mat 1_3\,,\quad\mat{C}=(B_\mathrm a+B_\mathrm b)\mat 1_3\,.
\end{equation}%
\begin{figure}[tp]
\centering
\includegraphics{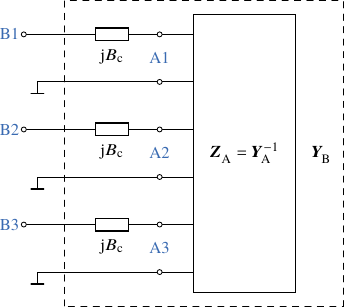}
\caption{Augmented antenna array.\label{fig:DMN_lehmb}}
\end{figure}%

To add a degree of freedom for easier design, we augment the admittance matrix of the antenna array as shown in Fig.~\ref{fig:DMN_lehmb} and get
\begin{equation}
\mat{Y}_{\!\mathrm B}=\left({\mat Y}_\mathrm{\!A}^{-1}+\frac{1}{\mathrm j B_\mathrm c }\mat 1_3\right)^{-1}
=\begin{bmatrix}e&c&c\\c&e&c\\c&c&e\end{bmatrix}
=\begin{bmatrix}\alpha&\beta&\beta\\\beta&\alpha&\beta\\\beta&\beta&\alpha\end{bmatrix}
\end{equation}
with $e=\chi-\mathrm j B_\mathrm c ^{-1}$.
The susceptance $B_\mathrm c$ is chosen to make the $\mathrm{Re}\{\mat Y_{\!\mathrm B}\}$ diagonal, because thsi allows us to decouple the ports P1, P2 and P3 by choosing
\begin{equation}
\mat C = \mat 0\,,\quad\mat A = -\mathrm{Im}\{\mat Y_\mathrm{\! \mathrm A}\}\label{eq:imagYa}\,,
\end{equation}
leading to
\begin{equation}
B_\mathrm t = \mathrm{Im}\{\beta\}\,,\label{eq:30}
\end{equation}
\begin{equation}
B_\mathrm b +B_\mathrm s = -\mathrm{Im}\{\alpha+2\beta\}\label{eq:31}\,,
\end{equation}
\begin{equation}
B_\mathrm a +B_\mathrm b = 0\,.\label{eq:32}
\end{equation}
For the admittance of the decoupled ports, we get
\begin{equation}
\mat Y = B_\mathrm b ^2\, \big(\mathrm{Re}\{\mat Y_\mathrm B\}\big)^{-1}=
\frac{B_\mathrm b^2}{\mathrm{Re}\{\alpha\}}\mat1_3
\end{equation}
with
\begin{equation}
B_b^{-2}\mathrm{Re}\{\alpha\}=Z_0\,,
\end{equation}
The condition
\begin{equation}
\mathrm{Re}\{\beta\}=
\mathrm{Re}\left\{\frac{c(c-e)}{e^3+2c^3-3c^2e}\right\}=0
\end{equation}
leads to a fourth-order equation for $e$, resulting in
\begin{equation}
B_\mathrm c=\frac{1}{\mathrm{Im}\{ \chi\}-\mathrm{Im}\{e\}}\,.
\end{equation}
From the four solutions for $e$, we choose a real-valued on which leas to the most convenient $B_\mathrm c$ value.
Finally, with the knowledge of $B_\mathrm c$, \eqref{eq:30}, \eqref{eq:31}, and \eqref{eq:32}, we have completed the design.

Next, an implementation with transmission lines is discussed. 
The matching step can be realized in many different (and well-known) fashions.
Later on, we employ transmission lines of suitable lengths and impedances instead of $B_\mathrm a$ and $B_\mathrm b$.

\begin{figure}[tp]
\centering
\includegraphics{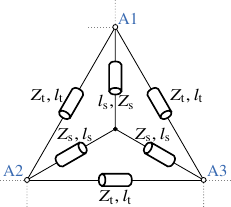}
\caption{Core part of star-triangle DMN realized with  transmission lines.\label{fig:DMN_lehm_tl}}
\end{figure}

Replacing the inner star-triangle part of Fig.~\ref{fig:DMN_lehma} (consisting of the floating susceptances $B_\mathrm t$ and the grounded susceptances $B_\mathrm s$) by the star-triangle of the transmission lines in Fig.~\ref{fig:DMN_lehm_tl}, 
we have to make sure that the admittance matrix $\mat A$ of the lumped circuit has to be equal to the admittance matrix
\begin{equation}
\mat Y_\mathrm{tl}=\begin{bmatrix}
 2Y_\mathrm t+Y_\mathrm s -Y_\mathrm s '&-Y_\mathrm t -Y_\mathrm s '&-Y_\mathrm t -Y_\mathrm s '\\
-Y_\mathrm t -Y_\mathrm s '& 2Y_\mathrm t+Y_\mathrm s -Y_\mathrm s '&-Y_\mathrm t -Y_\mathrm s '\\
-Y_\mathrm t -Y_\mathrm s '&-Y_\mathrm t -Y_\mathrm s '& 2Y_\mathrm t+Y_\mathrm s -Y_\mathrm s '
\end{bmatrix}
\end{equation}
 of the threeport depicted in Fig.~\ref{fig:DMN_lehm_tl}, with
\begin{equation}
Y_\mathrm t =\frac{-\mathrm j }{Z_\mathrm t \tan k _\mathrm t l_\mathrm t}\,,~~
Y_\mathrm s =\frac{-\mathrm j }{Z_\mathrm s \tan k _\mathrm s l_\mathrm s}\,,~~
Y_\mathrm s'=\frac{-2\mathrm j }{3Z_\mathrm s \sin 2k _\mathrm s l_\mathrm s}\,.
\end{equation}
We get a condition for the  diagonal matrix entries
\begin{equation}
2\mathrm j B_\mathrm t + \mathrm jB_\mathrm s=2Y_\mathrm t+Y_\mathrm s -Y_\mathrm s '
\label{eq:on}
\end{equation}
and one for the off-diagonal elements 
\begin{equation}
-\mathrm jB_\mathrm t = Y_\mathrm t -Y_\mathrm s '
\label{eq:off}\,.
\end{equation}

In addition to the four possible choices of $B_\mathrm c$, there is some design freedom hidden in \eqref{eq:on} and \eqref{eq:off}. 
For the two prescribed lumped element values, six design variables (two line impedance, two line lengths, and two wave numbers) are to be chosen.
However, this freedom is indeed necessary for any real-world design: 
The transmission line lengths
\begin{equation}
l_\mathrm t \ge d,\quad l_\mathrm s \approx d/\sqrt{3}=r_\textrm{uca}
\end{equation}
should be feasibly chosen as indicated by the antenna array element distance. 
For microstrip realizations as discussed later, the substrate material and height basically determine the propagation constants. 
Then, it is straightforward to calculate the line impedances from \eqref{eq:on} and \eqref{eq:off}.

It is worth noting that the circuit complexity of this DMN is reduced compared to the two-stage DMN with 33 pieces of transmission lines. 
In Fig.~\ref{fig:DMN_lehm_tl}, we have 6 pieces, where matching can be achieved by 3 additional serial lines. Furthermore, $l_s\approx r_\textrm{uca}$ can be much shorter than half a wavelength.
Compared to the decoupling-lines DMN in Fig.~\ref{fig:declin}(b), the star-triangle DMN exhibits additional decoupling paths by the star configuration. Implications of these additional design freedoms are revealed when discussing the design. 

\section{DMN Implementation, Fabrication, and S-Parameter Measurement} %

The simulations preceding the DMN realizations have been performed on an equivalent circuit level in Keysight Advanced Design System (ADS)~\cite{ADS} and as full-wave simulations in CST MWS. 
The design process involves several steps discussed in the following. 
Basically, this procedure was followed for each of the DMN designs.

The two starting ingredients for the initial ADS simulations are 
i) S-parameter data sets obtained by full-wave CST MWS simulations of the UCA with microstrip feed: These data sets comprise parameter variations of the UCA distance $d$ (with individually optimized monopole length) and of the coax-microstrip transition.
ii) Furthermore, the ideal equivalent circuit of the DMN under consideration: 
First, the DMN is constructed with ideal elements in ADS and initial parameter values are calculated according to the analytical derivations in Section~\ref{sc3}. 
This DMN of course shows excellent decoupling and matching at the single design frequency.
Subsequently, optimizations for a wider bandwidth are carried out based on microstrip transmission line models. 
Different parameters of the UCA implementation are studied. 

At this point, practical implications begin to play an important role: 
Some of the transmission line lengths are heavily restricted by the dimensions of the UCA.
For instance, the minimum length of lines connecting two antenna elements is given by the distance between the antennas. The maximum length should not be much larger. 
The line impedances should also be restricted to reasonable choices of the microstrip line width. 
Hence, the line length and width variables are optimized in ADS for a feasible solution; simultaneously, broadband matching and decoupling goals are considered. 
This procedure constitutes optimization cycles executed until all microstrip lengths and widths seem feasible and the bandwidths are satisfyingly large. 
Going back to the first step, i.e., choosing a UCA with different dimensions, is still possible if no satisfying solution is found. 

Eventually, the microstrip circuit is implemented in CST MWS and further full-wave optimizations of all parameters are carried out. 
On the one hand, this is necessary since  a simultaneous UCA optimization is not possible on a circuit level; 
on the other hand, the circuit functionality is changed to a small extent since some effects (e.g., the behavior of microstrip junctions, bends, and line-to-line coupling) are not fully included in ADS. 
The tuned parameters include the antenna parameters such as array spacing or monopole lengths together with all DMN parameters. 
Overall, only very few design parameters were fixed, see the setup description in Section IIB. The goal was to achieve the best possible UCA/DMN performance for each DMN approach.

\subsection{Design of the Neutralization-Lines DMN}

\begin{figure}[tp]
\centering
\includegraphics[]{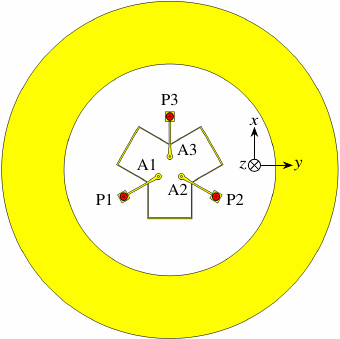}
\caption{CST MWS simulation model of neutralization-lines DMN, rear side.\label{fig:declin_CST}}
\end{figure}

This DMN exhibits a quite simple structure, see the circuit model in Fig.~\ref{fig:declin}(b) and the final design in Fig.~\ref{fig:declin_CST}.
The radius is $r_\textrm{uca}=8.6\,\mathrm{mm}=0.103\lambda_0$, which was subject to the optimization.
The design process is easily feasible; however, it turns out that the bandwidth is limited. 
The widths and lengths of all microstrip lines are attained via the described optimization procedure of subsequent ADS and CST MWS simulations.

\begin{figure}[tp]
\centering
\includegraphics[]{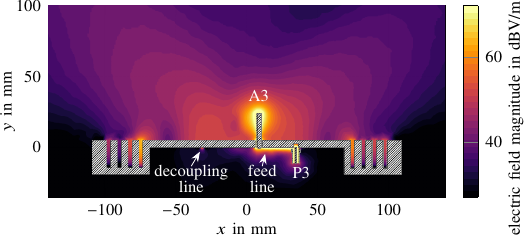}
\caption{Electric near field evaluations for neutralization-lines DMN.\label{fig:nf}}
\end{figure}

For this particular array, the circular groundplane was modified with quarter-wavelength grooves at its edge to support the wave detachment  and to suppress radiation behind the ground plane caused by edge diffraction. This is visualized in Fig.~\ref{fig:nf} and results in a larger realized gain above the groundplane, i.e., in $+z$-direction.
Due to the corrugated groundplane, single-element radiation patterns and gains of the array are only qualitatively comparable to the other arrays since the corrugated groundplane causes an increase in gain. Nevertheless, the corrugations are an interesting means for modifying the radiation behavior of the array, which can be useful for many applications.

\begin{figure}[tp]
\centering%
\vspace*{-0.2cm}%
\hspace*{0.2cm}%
\subfloat[]{\includegraphics[height=0.4\linewidth]{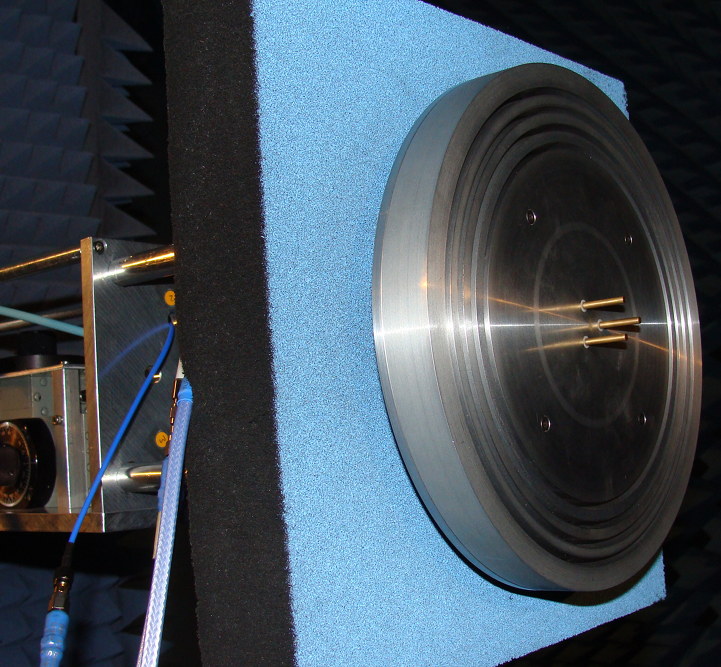}}%
\hfill%
\subfloat[]{\includegraphics[height=0.4\linewidth]{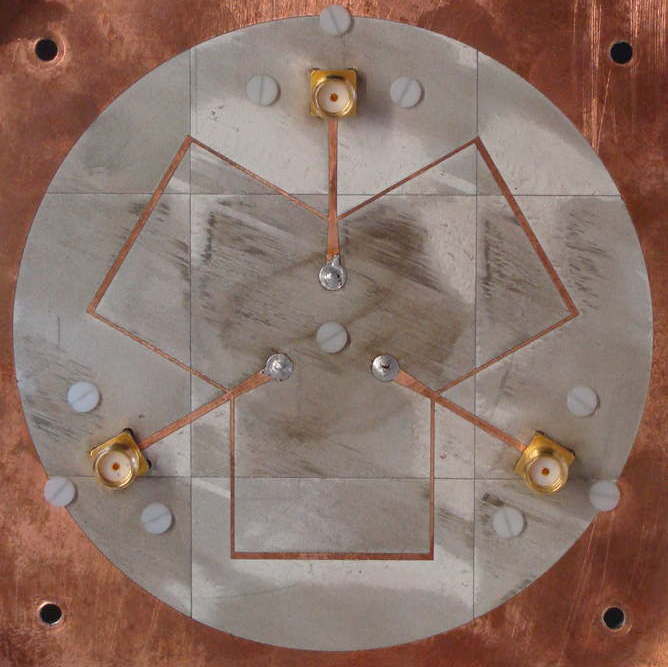}}%
\hspace*{0.2cm}%
\caption{Photographs of the 3-element UCA with neutralization-lines DMN and corrugated ground plane. (a)~Mounted in anechoic chamber, front view. (b)~PCB of DMN, rear view.\label{fig:DMN123}}
\end{figure}

\begin{figure}[!tp]
\centering
\includegraphics[]{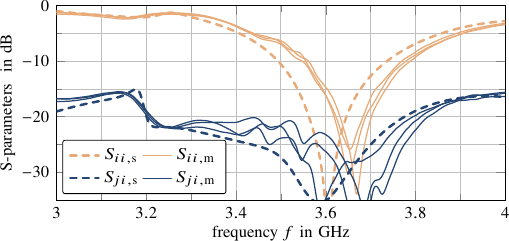}
\caption{Simulated scattering parameters $S_\mathrm s$ and measured ones $S_\mathrm m$ of three-element UCA with corrugated groundplane and neutralization-lines DMN.\label{fig:sparam_3_corr}}
\end{figure}

In Fig.~\ref{fig:DMN123}, photographs of the fabricated prototype show the DMN microstrip circuit on the rear side and the monopole UCA and the corrugated ground plane on the front side. 
The scattering parameters in Fig.~\ref{fig:sparam_3_corr} demonstrate a good agreement between measurement and full-wave simulation (neglecting the minimal frequency shift), and the $-16$\,dB bandwidth is about $100-110$\,MHz in the measurement and 120\,MHz in the simulation, limited only by the input matching due to the excellent decoupling.
To obtain these simulation results, a little adjustment had to be applied: The corrugations, which behave kind of resonant, have been filled with slightly lossy material of $\varepsilon_\mathrm r=1$ and $\tan\delta=0.01$ due to numerical problems in the frequency-domain solver of CST MWS. 
These additional losses influence the gain by about $-0.3$\,dB, which is corrected for in the pattern and gain discussion for a valid comparison of measurements and simulation.

\subsection{Design of the Two-Stage DMN}

During the discussed optimization procedure, it became clear that this two-stage DMN circuit allows a realization with a larger bandwidth  since it is not completely symmetric; it exhibits rotational symmetry only. 
As seen for instance in Fig.~\ref{fig_proposed}, the feed port P1 is connected to  the two antennas A1 and A3  via  separate susceptances $B_2$ and $B_3$. 
The ADS-optimized design was realized in a way that, for a lower frequency range, one antenna (A1 or A3) is matched and the two others are only excited for port decoupling purposes of P2 and P3.
For the upper part of the bandwidth, the other antenna (A3 or A1) is then matched and, again, the other ports P2 and P3 are decoupled. 
Over the whole bandwidth, a fluent transition in the matching and later on also in the far-field patterns is observed.
In other words, the antenna A1 is matched to either P2 or P1 dependent on frequency.

\begin{figure}[tp]
\centering%
\vspace*{-0.2cm}
\subfloat[]{\includegraphics[]{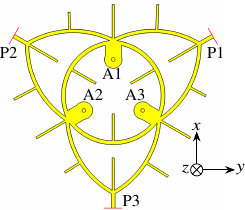}}%
\hfill%
\subfloat[]{\includegraphics[height=0.4\linewidth]{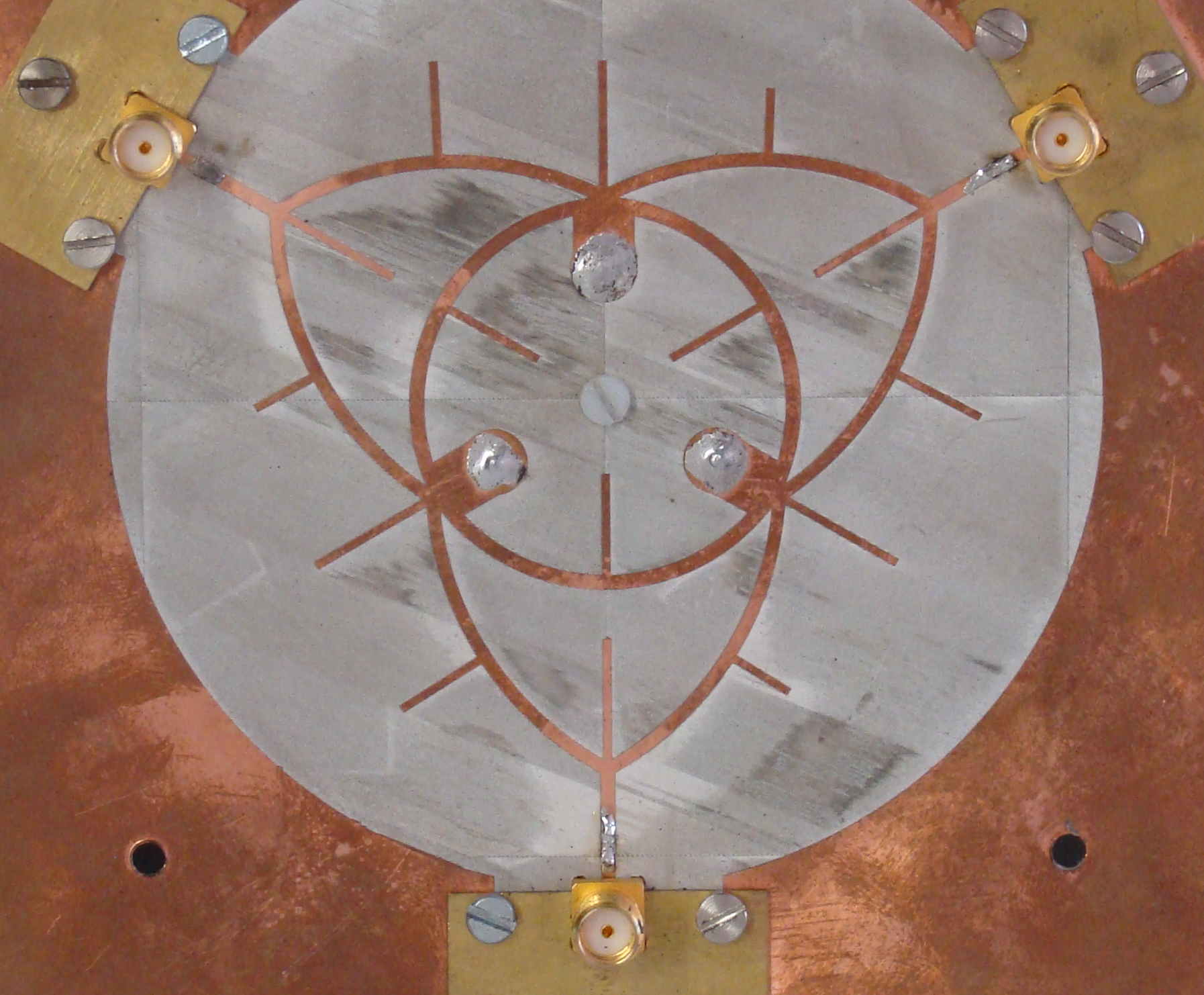}}%
\caption{The implemented version of the two-stage DMN connected to a three-element UCA. (a)~Simulation model in CST MWS. (b)~Photograph of fabricated PCB.\label{fig:DMNts_photo}}
\end{figure}

The CST MWS simulation model in Fig.~\ref{fig:DMNts_photo} adapts the circuit of Fig.~\ref{fig:DMN_ts_tl}. 
The transmission lines can of course not be straight lines but have to be bent to fulfill the constraints given by the UCA with $r_\mathrm{uca}=11.08\,\mathrm{mm}=0.136\lambda_0$. 
The major difference to Fig.~\ref{fig:DMN_ts_tl} is that there are (quite broad) microstrip lines from the antenna ports A1--A3 to the entry point of the discussed DMN; introducing this additional antenna feed line allows for more freedom in the DMN/UCA optimization process and leads to a better solution.

\begin{figure}[tp]
\centering
\includegraphics[]{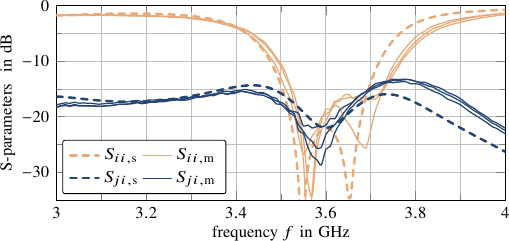}
\caption{Scattering parameters of three-element UCA with two-stage DMN, simulated is $S_\mathrm s$ and measured $S_\mathrm m$.\label{fig:sparam_3_ts}}
\end{figure}

In Fig.~\ref{fig:sparam_3_ts}, 
the measured and simulated S-parameters agree very well. The $-16$-dB matching bandwidth is 185\,MHz in simulation and almost $200$\,MHz in measurement. The coupling stays below the $-16$-dB threshold above $3.5$\,GHz in simulation and shows a bandwidth of over $200$\,MHz in the measurement.
This DMN can achieve approximately twice the bandwidth of the neutralization-lines DMN since it exhibits more design freedom.

\subsection{Design of the Star-Triangle DMN}

\begin{figure}[tp]
\centering%
\vspace*{-0.2cm}
\subfloat[]{\includegraphics[]{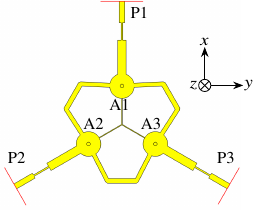}}%
\hfill%
\subfloat[]{\includegraphics[height=0.4\linewidth]{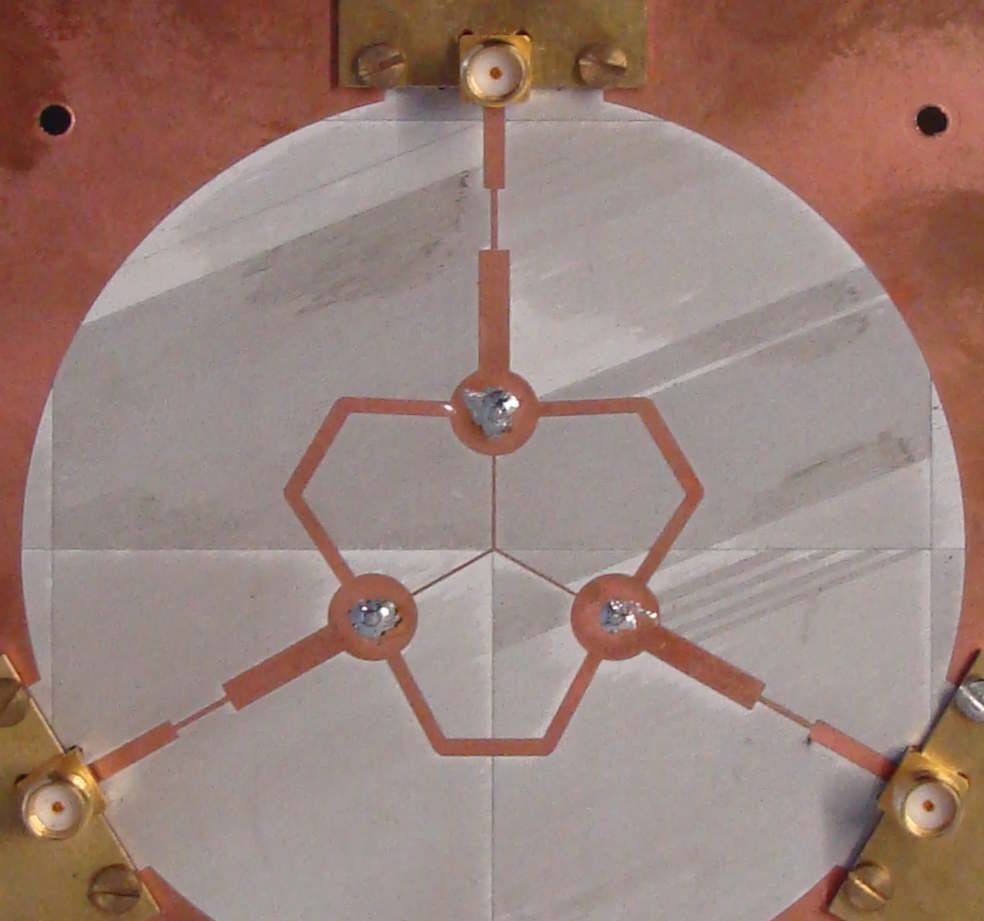}}%
\caption{The implemented version of the star-triangle DMN connected to a three-element UCA. (a)~Simulation model in CST MWS. (b)~Photograph of fabricated PCB.\label{fig:DMNst_photo}}
\end{figure}

In the same way as for the other two designs, the original single frequency-design was first optimized in ADS for a broadband matching and decoupling and afterwards modeled in CST MWS. This optimized structure is shown in Fig.~\ref{fig:DMNst_photo}.
There is one design difference to all other discussed DMNs: The substrate thickness $t_\mathrm s = 0.762\,\mathrm{mm}$  was chosen instead of $0.508$\,mm to minimize loss; this means a trade-off between dielectric and microstrip radiation loss.
Other design choices include the UCA radius  $r_\mathrm{uca}=12.7\,\mathrm{mm}=0.152\lambda_0$, the matching with two cascaded transmission lines at the port instead of lumped susceptances and an optimized antenna connection instead of a parallel susceptance at the antenna feed points.

\begin{figure}[tp]
\centering
\includegraphics[]{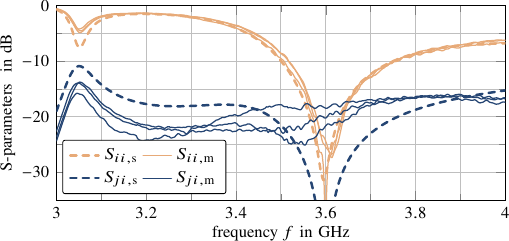}
\caption{Simulated scattering parameters $S_\mathrm s$ and measured ones $S_\mathrm m$ of three-element UCA with star-triangle DMN.\label{fig:sparam_3_st}}
\end{figure}

In Fig.~\ref{fig:sparam_3_st},  the S-parameters are compared, where the decoupling is below $-16$\,dB for a very broad bandwidth and the $-16$-dB input impedance bandwidth is 120\,MHz in simulation and $105-110$\,MHz in the measurement.

\subsection{Total Efficiencies of Decoupled Arrays}

\begin{figure}[!tp]
\centering
\includegraphics[]{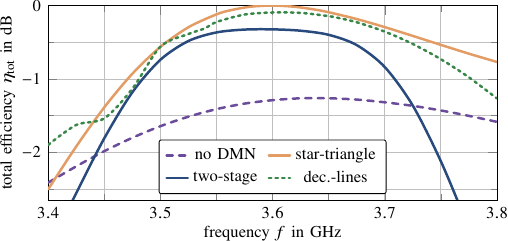}
\caption{Simulated total efficiencies  of the realized DMNs.\label{fig:effs}}
\end{figure}

We consider the total efficiencies $\eta_\mathrm{tot}$ in Fig.~\ref{fig:effs}, including loss and mismatch.
Every DMN offers a better efficiency than the UCA without DMN between $3.5$ to $3.7$\,GHz. 
Interesting to note is that the star-triangle DMN offers the best total efficiency at any frequency even given that its S-parameters bandwidth was the same as for the neutralization-lines DMN and much smaller than the bandwidth of the two-stage DMN. Furthermore, it becomes obvious that the larger operating bandwidth of the two-stage DMN with respect to input matching is bought by a larger loss in its complicated microstrip structure. Analyzing the per-material loss in CST MWS reveals that the worse total efficiency of the two-stage DMN solely originates from around 80\si{\percent} copper conductivity loss and 20\si{\percent} dielectric substrate loss.

In a real scenario, one would choose the design according to the importance of design goals: 
If loss should be kept as low as possible, the star-triangle DMN offers the better solution; if wide impedance bandwidth is important, the two-stage DMN is preferable.

\subsection{A Four-Element Neutralization-Lines DMN}

\begin{figure}[tp]
\centering\vspace*{-0.2cm}
\settoheight{\imageheight}{\includegraphics[]{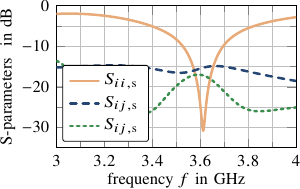}}
\subfloat[]{\tikz\node[minimum height=\imageheight]{\includegraphics[width=0.38\linewidth]{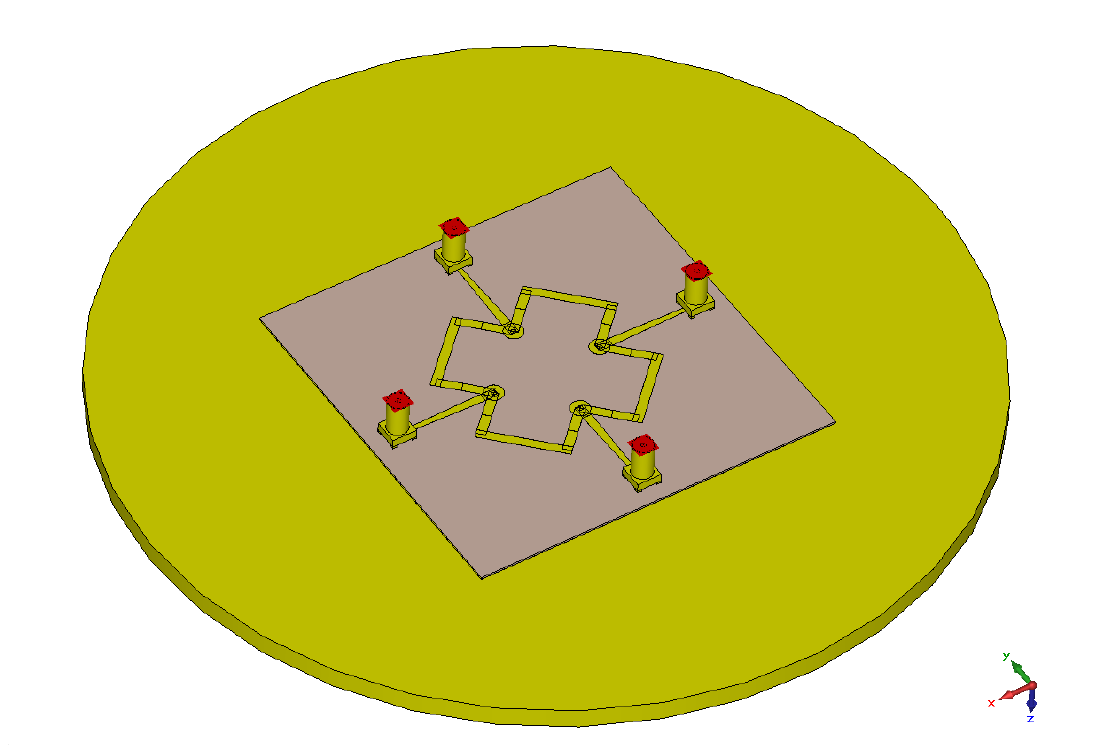}};}%
\hfill%
\subfloat[]{\includegraphics[]{fig26b}}%
\caption{Four-element array with a DMN consisting of simple neutralization lines. (a)~Simulation model. (b)~Simulated S-parameters.\label{fig:DMN_4}}
\end{figure}

For the evaluation of the differences between even- and odd-numbered UCAs, a DMN with neutralization lines was realized in simulation only for a four-element UCA. 
The radius  was adjusted to $r_\mathrm{uca}=0.16\lambda_0$. 
The DMN and its S-parameters are shown in Fig.~\ref{fig:DMN_4}.
Since the UCA radius is larger than for the four-element model in Section IID, the coupling without DMN is a bit lower and goes up to $-9$\,dB. With the DMN, this is reduced to maximally $-15$\,dB. 
The total efficiency varies between $-0.1$\,dB and $-0.5$\,dB, dependent on the beamsteering direction.

\begin{figure}[!tp]
\centering\vspace*{-0.3cm}
\subfloat[]{\includegraphics[]{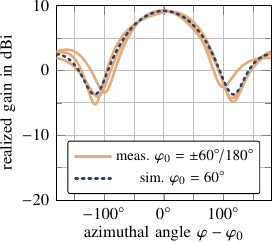}}%
\hfill%
\subfloat[]{\includegraphics[]{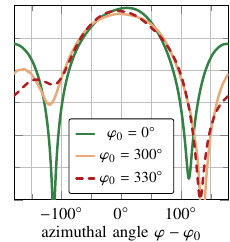}}%
\\%
\subfloat[]{\includegraphics[]{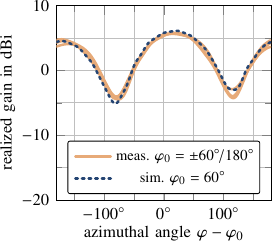}}%
\hfill%
\subfloat[]{\includegraphics[]{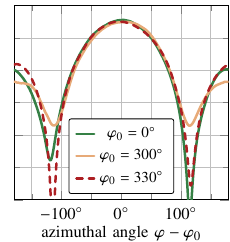}}%
\\%
\subfloat[]{\includegraphics[]{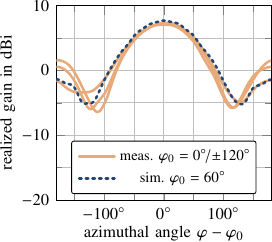}}%
\hfill%
\subfloat[]{\includegraphics[]{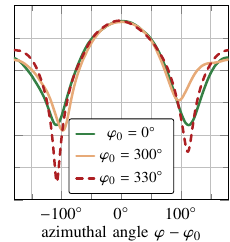}}%
\caption{Comparison of simulated and measured single excitation (i.e., active element) and array excitation realized gain patterns. (a)~Neutralization-lines DMN, single element patterns,  $\vartheta=40^\circ$ cut. (b)~Neutralization-lines DMN, array gain,  $\vartheta=40^\circ$ cut.
(c)~Two-stage DMN, single  element patterns, $\vartheta=70^\circ$. (d)~Two-stage DMN, measured patterns for optimized $\varphi_0$ beamsteering, $\vartheta=70^\circ$. (e)~Star-triangle DMN, single  element patterns, $\vartheta=70^\circ$. (f)~Star-triangle DMN, measured patterns for optimized $\varphi_0$ beamsteering, $\vartheta=70^\circ$. 
\label{fig:final_cuts}}
\end{figure}%

\section{Gain Measurements For Single-Port Excitation and Beamforming Capability Verification} %

\subsection{Far-Field Patterns}

Realized gain measurements of the three-element UCAs with DMN have been performed in the anechoic chamber of TUM. 
The measurement results of individual port patterns and beamformed patterns are summarized in Fig.~\ref{fig:final_cuts}.

Due to the corrugations at the ground plane edge, the pattern of the UCA with neutralization-lines DMN has the maximum at around $\vartheta=40^\circ$, not at $\vartheta=70^\circ$ as all other UCAs.
Hence, the individual-port patterns are given for this $40^\circ$-cut in Fig.~\ref{fig:final_cuts}(a), where an increase in gain is observed due to the suppressed radiation behind the ground plane. Furthermore, this radiation pattern is evaluated at $3.65$\,GHz, the frequency with the best matching in the measured S-parameters. 
The measurements agree quite well with the simulation  around the main beam. 
The beamforming weights for the analogously-fed and measured optimum excitation coefficients have been computed in the $\vartheta=30^\circ$ cut; the beamsteering effect is nevertheless obvious in Fig.~\ref{fig:final_cuts}(b) for the main-beam directions $\varphi_0=0^\circ$, $\varphi_0=300^\circ$, and $\varphi_0=330^\circ$.

The same comparisons of measured and simulated single-excitation patterns are performed for the two-stage DMN in Fig.~\ref{fig:final_cuts}(c) and the comparison of the respective beamsteered gain pattern measurements are shown in Fig.~\ref{fig:final_cuts}(d). Both measurements are analyzed in the $\vartheta=70^\circ$ cut.
The lower gain values of this UCA are expected due to the corrugated ground plane in the first UCA. 
The agreement with the simulated pattern and the beamsteering capabilities are well confirmed.

The very same comparisons are shown for the the star-triangle DMN in Fig.~\ref{fig:final_cuts}(e) and (f). The gain values of the simulation and the beamforming behavior are demonstrated.

\subsection{Azimuthal Beamforming}

Section II focused on even- and odd-numbered UCAs: 
In a theoretical investigation based on array factors, we postulated that the achievable gain of odd-numbered UCAs is rather constant over the beamsteering angle, whereas even-numbered UCAs with one element more show the same maximally possible gain in certain directions and a larger gain in between.
Another insight was that DMNs are needed to realize these effects and achieve the maximum beamforming performance.

\begin{figure}[tp]
\centering
\includegraphics[]{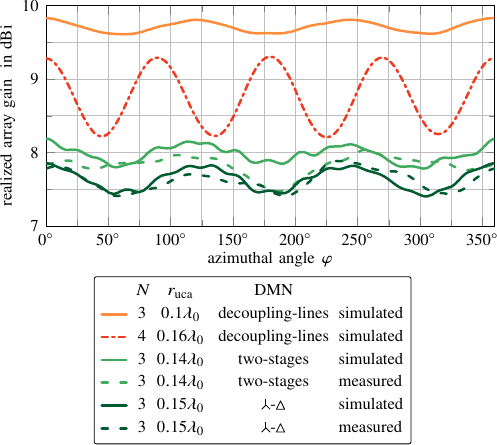}
\caption{Optimum array gain dependent on the scan angle for simulations and measurements of three- and four-element UCAs with DMNs. The three-element neutralization-lines DMN is evaluated at $\vartheta=40^\circ$, all others at $\vartheta=70^\circ$.\label{fig:gainreal2}}
\end{figure}

We have in mind that the gain in the $\vartheta=70^\circ$ cut for both three- and four-element UCAs was about $6.7$\,dBi on average for azimuthal beamsteering, see Fig.~\ref{fig:gainreal}.
Similar gain curves are presented in Fig.~\ref{fig:gainreal2} 
for arrays with the two-stage and star-triangle DMNs as well as for the array with the four-element neutralization-lines DMN. 
The beamforming results in Fig.~\ref{fig:gainreal2} are based on array realized gain patterns for each beamsteering angle, as shown for exemplary angles in Fig.~\ref{fig:final_cuts}.
We want to emphasize two findings.
First, the gain fluctuations of the four-element UCA with DMN are larger than for the three-element UCAs, confirming the investigations in Section II.
Second, the star-triangle DMN indeed achieves the same array gain as the minimum gain value of the four-element UCA with DMN.
The two-stage DMN exhibits a bit lower gain due to its worse total efficiency.

The neutralization-lines DMN with the corrugated ground plane is also included in Fig.~\ref{fig:gainreal2}, again for $\vartheta=40^\circ$ as a special case. Since the radiation behind the groundplane is effectively suppressed, the gain of almost 9\,dBi comes very close to the ideally expected values of almost $10$\,dBi seen in in Figs.~\ref{fig:ag1} and~\ref{fig:gain_PM}.

Finally, we summarize the performance of all three-element arrays and their different DMNs in Tab.~\ref{tab:final}, which is based on simulation results.
The proposed array configurations are compared regarding relative $-16$\,dB bandwidth \textit{BW}, efficiency, realized gain, UCA radius, and number of radiators to designs from literature.
\begin{table}[tp]
\newcommand{\fc}{$f_\mathrm{c}\sfrac{}{\mathrm{GHz}}$}
\newcommand{\BW}{$\mathit{BW}$}
\newcommand{\pc}[1]{$#1\si{\percent}$}
\newcommand{\ntot}{$\eta_\mathrm{tot}\sfrac{}{\mathrm{dB}}$}
\newcommand{\Sji}{$S_{ji}\sfrac{}{\mathrm{dB}}$}
\newcommand{\Greal}{$G\sfrac{}{\mathrm{dBi}}$}
\caption{Comparison of UCA/DMN properties to corresponding designs from literature regarding center frequency $f_c$, relative $-16$\,\textup{d}B bandwidth $\textit{BW}$ in relation to $f_c$, maximum coupling $S_{ji}$ withing the bandwidth, maximum total efficiency $\eta_\mathrm{tot}$, realized gain $G$, UCA radius $r_\mathrm{uca}$, and number of UCA elements $N$.\label{tab:final}}
\centering
\def\startt{\tikz{\draw[cap=round] (0,0) -- (0,0.096cm);\draw[cap=round] (0,0) -- (0.07cm,-0.07cm);\draw[cap=round] (0,0) -- (-0.07cm,-0.07cm);}-\tikz{\draw[cap=round,rounded corners=0pt] (0,0) -- (0.14cm,0) -- (0.07cm,0.14cm) -- cycle;}}
\begin{tabular}{lrrrrrrr}
\toprule[1pt]
                  &    \fc &     \BW  & \Sji &         \ntot   &\Greal & $r_\mathrm{uca}\sfrac{}{\lambda_0}$ & $N$ \\
 \cmidrule[0.5pt](lr){2-2} \cmidrule[0.5pt](lr){3-3} \cmidrule[0.5pt](lr){4-4} \cmidrule[0.5pt](lr){5-5}\cmidrule[0.5pt](lr){6-6}\cmidrule[0.5pt](lr){7-7}\cmidrule[0.5pt](lr){8-8}
no DMN            & $ 3.6$ &     ---  & $ - 9$ &         $-1.28$ & $6.8$ &   $0.1$ & $3$ \\
dec.-lines        & $ 3.6$ & \pc{3.3} & $<-20$ &          $-0.09$ & $9.7$ &   $0.1$ & $3$ \\  
two-stage         & $ 3.6$ & \pc{5.1} & $ -16$ &          $-0.32$ & $7.6$ &  $0.14$ & $3$ \\  
\startt           & $ 3.6$ & \pc{3.3} & $<-20$ &  $\approx\,0$    & $8.0$ &  $0.15$ & $3$ \\  
\cite{Wang2013}   & $2.45$ & \pc{2.9} & $ -18$ &          $-0.96$ & $5.5$ &  $0.05$ & $2$ \\  
\cite{Wang2013}   & $2.55$ & \pc{1.5} & $<-20$ &          $-3.28$ & $3.0$ & $0.041$ & $4$ \\  
\cite{li2019novel}& $ 2.4$ & \pc{1.0} & $ -16$ &          $-0.46$ & $4.9$ & $0.035$ & $2$ \\  
\cite{li2019novel}& $ 2.4$ &     ---  & $<-20$ &          $ -0.9$ &  ---  & $0.246$ & $3$ \\ 
\bottomrule[1pt]
\end{tabular}
\end{table}
The $-16$\,dB matching bandwidths and the decoupling performances within this bandwidth have already been discussed in detail. 
The total efficiencies at the design frequency reveal that there is an improvement of $-1$\,dB to $-1.3$\,dB.
This is reflected in the realized-gain increase, too, where the averaged realized gain $G_\mathrm{real}$ in the cut of Fig.~\ref{fig:gainreal2} is considered. 
The UCA with neutralization-lines DMN, which shows a gain of $9.7$\,dBi in the $\vartheta=40^\circ$ cut, is a special case. At $\vartheta=70^\circ$, the gain drops to only $5.1$\,dBi.
In the comparison in Tab.~\ref{tab:final}, we have to keep in mind that each of our proposed designs (DMN plus UCA) was optimized for the its best performance as a system. 
Here, it was in particular found that a varying element distance helped to improve the system performance for some of the designs. Therefore, our comparison relates to properties of the systems consisting of DMN and UCA and it should not just be seen as a comparison of UCAs with a certain geometric arrangement.

\section{Conclusion} %

Two new DMN circuits for three-element UCAs have been introduced, analyzed, designed, fabricated, and verified in measurements. 
The proposed DMNs were analyzed and design formulas for arbitrary antenna impedances have been provided. In the practical validation, simple monopole radiators have been utilized to realize compact UCAs.
The DMNs were built in microstrip technology. 
All of them are realized without cross-overs which is very convenient for planar PCBs.
In this respect, it was demonstrated that odd-numbered UCAs are capable of offering an approximately constant gain in azimuthal cuts whereas even-numbered ones (with one element more) show gain fluctuations, where the minimum has about the same gain as odd-numbered UCAs. We conclude that in general the same array performance is achieved  with odd-numbered UCAs, which was investigated in detail for arrays with three and four elements.
This was investigated in theory, in simulation, and by measurement.

In addition to the theoretical insights, one of the presented DMNs provides an excellent total efficiency over a large bandwidth, while the other one exhibits an excellent matching and decoupling bandwidth bought by reduced efficiency.
As compared to plain UCAs without DMNs, the decoupled and matched UCAs offer a gain and efficiency improvement of about $1$ to $1.5$\,dB.
These improvements have been achieved at the expense of more complex DMNs, which have a relatively large size in our realizations. 
However, compact realizations of the DMNs with lumped elements (as discussed theoretically in the paper), with transmission lines in multilayer PCBs, or even with shielded transmission lines are also possible.

\bibliographystyle{IEEEtran}
\bibliography{IEEEabrv,ref}

\begin{IEEEbiography}
 [{\includegraphics[width=1in,height=1.25in,clip,keepaspectratio]{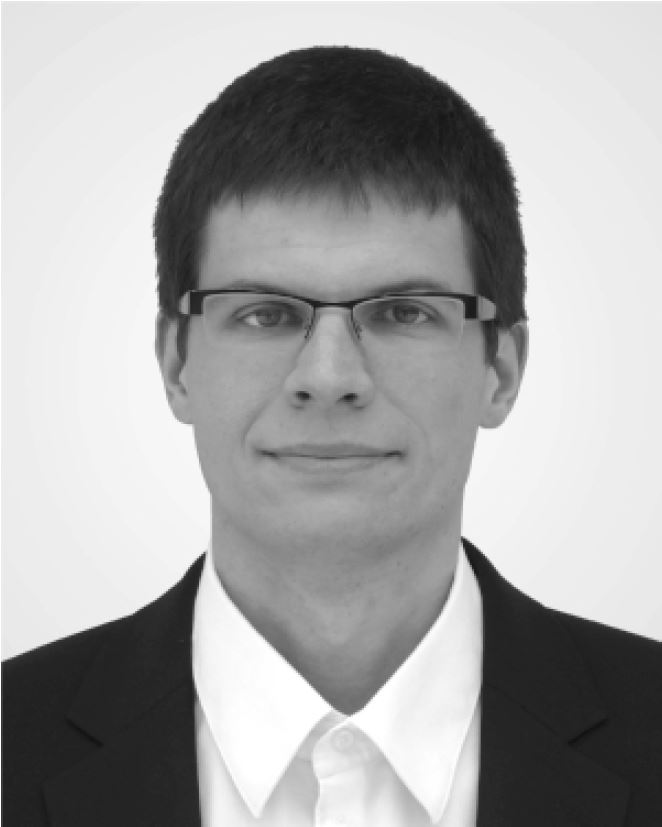}}]
{Jonas Kornprobst} (S’17) received the B.Eng. degree in electrical engineering and information technology from the University of Applied Sciences Rosenheim, Rosenheim, Germany, in 2014, and the M.Sc. degree in electrical engineering and information technology from the Technical University of	Munich, Munich, Germany, in 2016. 

Since 2016, he has been a Research Assistant with the Chair of High-Frequency Engineering, Department of Electrical and Computer Engineering, Technical University of Munich. 
His current research interests include numerical electromagnetics, in particular integral equation	methods, antenna measurement techniques, antenna and antenna array design, as well as microwave circuits.
\end{IEEEbiography}
\begin{IEEEbiography}
 [{\includegraphics[width=1in,height=1.25in,clip,keepaspectratio]{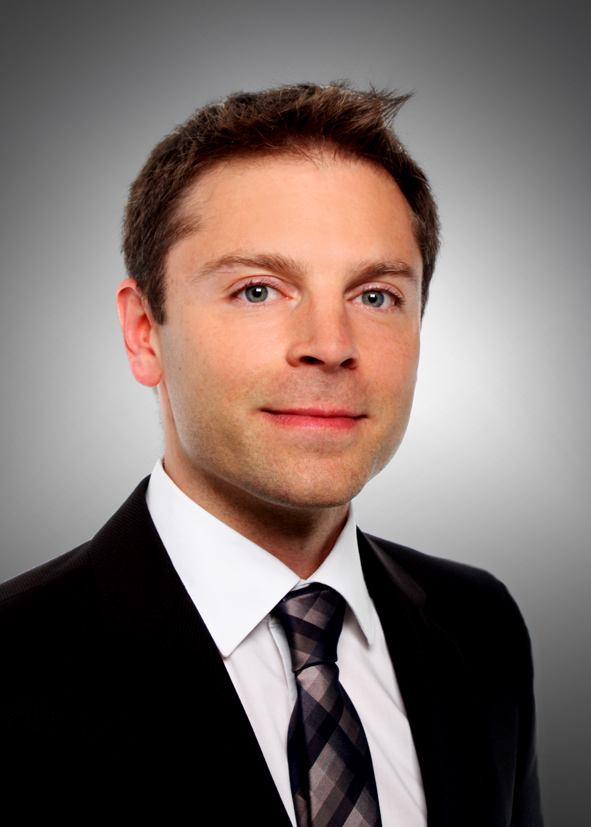}}]
{Thomas J. Mittermaier} (S'12$-$M'17) received the Dipl.-Ing.(FH) degree from Fachhochschule Regensburg, Regensburg, Germany, and the M.Sc. degree from the Technical University of Munich, Munich, Germany, in 2009 and 2011, respectively, both in electrical engineering and information technology. 

From 2011 to 2017 he was a Research Assistant at the Chair of High-Frequency Engineering of the Technical University of Munich, Munich, Germany. 
His doctoral research focused on signal processing and tracking algorithms for extremely short-range radar in industrial machine tools as well as microwave circuits. In May 2017, he joined the RF360 Europe GmbH, Munich, as a Development
Engineer for bulk acoustic wave devices. 
He is currently involved in linear and nonlinear modeling, in power durability and life-time testing of acoustic resonators and filters for future mobile communications.
\end{IEEEbiography}
\begin{IEEEbiography}
 [{\includegraphics[width=1in,height=1.25in,clip,keepaspectratio]{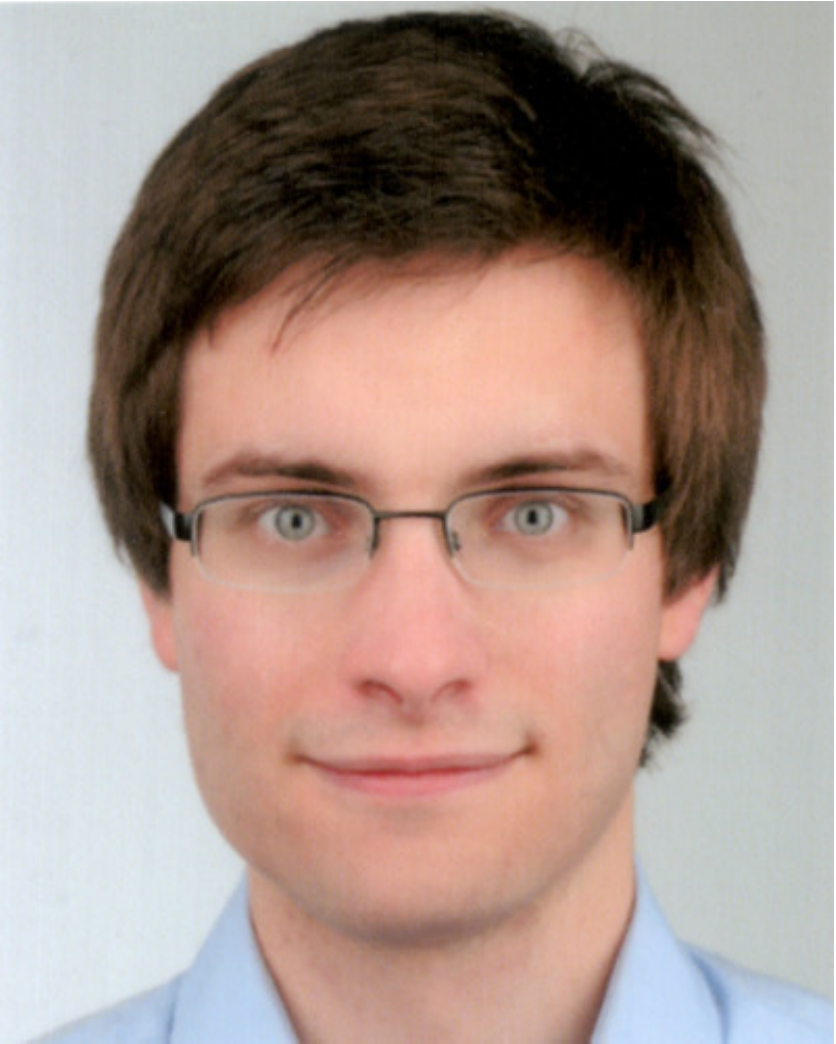}}]
{Raimund A. M. Mauermayer} (S'14) received the Dipl.-Ing. degree in electrical engineering and information technology from the Technical University of Munich, Munich, Germany, in 2012.

From 2012 to 2018, he was a Research Assistant with the Chair of High-Frequency Engineering, Technical University of Munich. 
In 2019, he joined the Department of Antenna Design and Measurement, Mercedes-Benz AG, Sindelfingen, Germany. 
His research interests are near-field far-field transformation and antenna measurement techniques.
\end{IEEEbiography}
\begin{IEEEbiography}
 [{\includegraphics[width=1in,height=1.25in,clip,keepaspectratio]{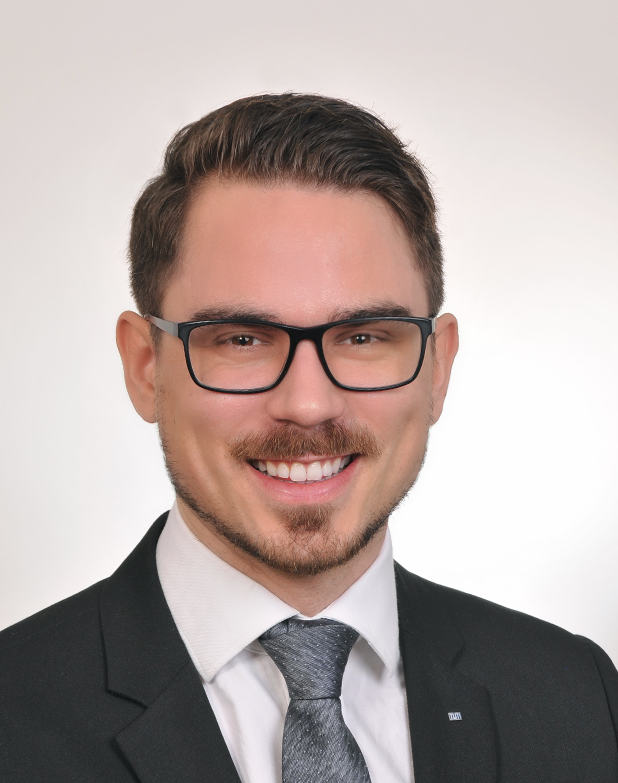}}]
{Matthias G. Ehrnsperger} (S'19) received the B.Eng. degree in electrical engineering and information technology from the Technical University of Applied Sciences Regensburg, Regensburg, Germany, in 2015, and the M.Sc. degree in electrical engineering and information technology from the Technical University of Munich, Munich, Germany, in 2017.

Since 2017, he has been a Research Assistant with the Chair of High-Frequency Engineering, Department of Electrical and Computer Engineering, Technical University of Munich. 
His current research interests include radar systems, in particular antenna and antenna array design, rapid prototyping methodologies, statistical- and machine learning based signal processing, as well as microwave circuits.
\end{IEEEbiography}
\begin{IEEEbiography}
 [{\includegraphics[width=1in,height=1.25in,clip,keepaspectratio]{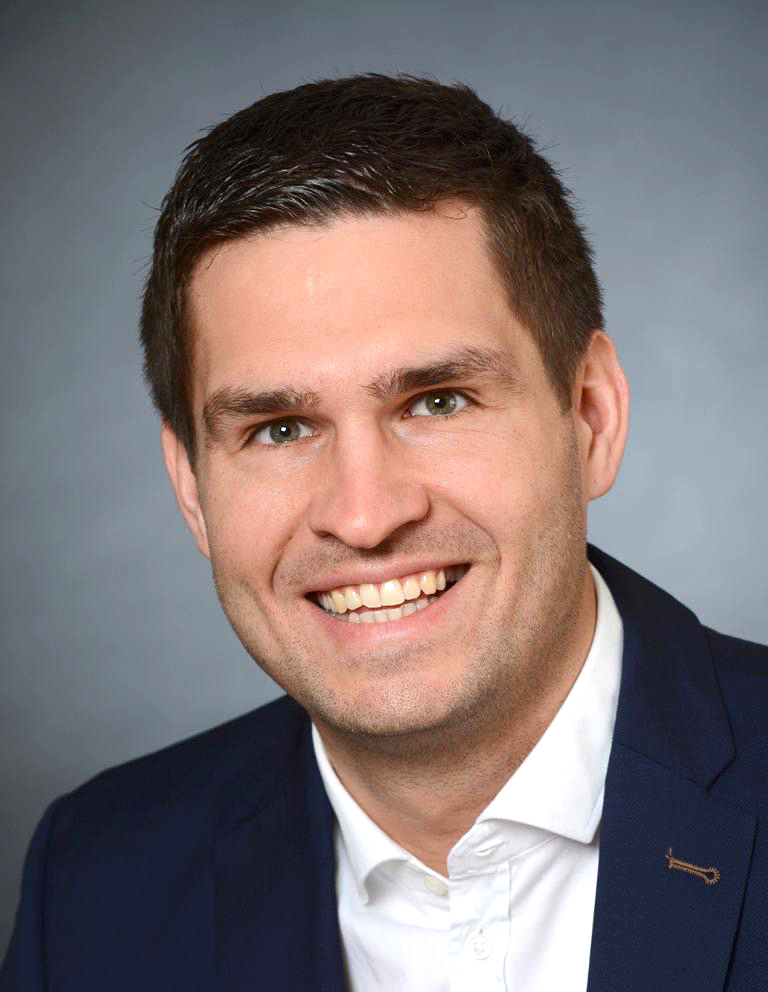}}]
{Gerhard F. Hamberger} received the B.Sc., the M.Sc., and the Dr.-Ing. degree in electrical and computer engineering from the Technical University of Munich, Munich, Germany, in 2012, 2014, and 2019, respectively.

In 2018, he had joined the Systems and Projects Department of Rohde \& Schwarz, Munich, Germany, where he was working on various state-of-the-art 5G antenna and device test systems and field transformation software. In 2019, Dr.-Ing. Hamberger has become a Senior Development Engineer in the Microwave Imaging Department of Rohde \& Schwarz, Munich, Germany, focusing on automotive radar and radome test systems
\end{IEEEbiography}
\begin{IEEEbiography}
 [{\includegraphics[width=1in,height=1.25in,clip,keepaspectratio]{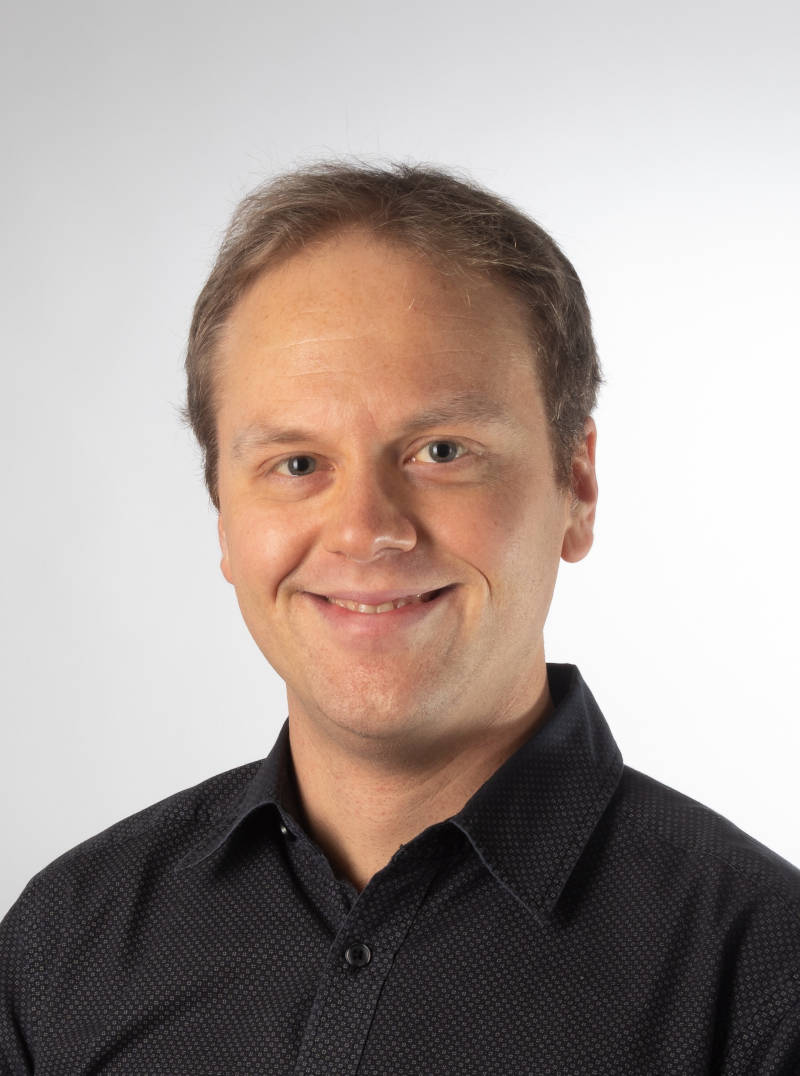}}]
{Bernhard Lehmeyer} received the Dipl.-Ing. and Dr.-Ing. degrees in electrical and computer engineering from the Technical University of Munich (TUM), Munich, Germany, in 2010 and 2018.

In 2010, he joined Rohde \& Schwarz GmbH \& Co. KG Germany as circuit design engineer. 
From 2012 to 2018, he was a research assistant at the institute of circuit theory and signal processing at the Technical University of Munich. 
From 2018 to 2019, he was a postdoctoral researcher at the High Voltage Lab/ETHZ, Swiss Federal Institute of Technology in Zurich, Zurich, Switzerland. 
Since 2019, he is the CEO of Tronika GmbH Germany. 
His current research interests are RF circuitry, optical detector design and power inverter optimization.
\end{IEEEbiography}
\begin{IEEEbiographynophoto}
{Michel T. Ivrla\v{c},} 	biography not available at time of publication.
\end{IEEEbiographynophoto}
\vfill
\newpage
\begin{IEEEbiography}
 [{\includegraphics[width=1in,height=1.25in,clip,keepaspectratio]{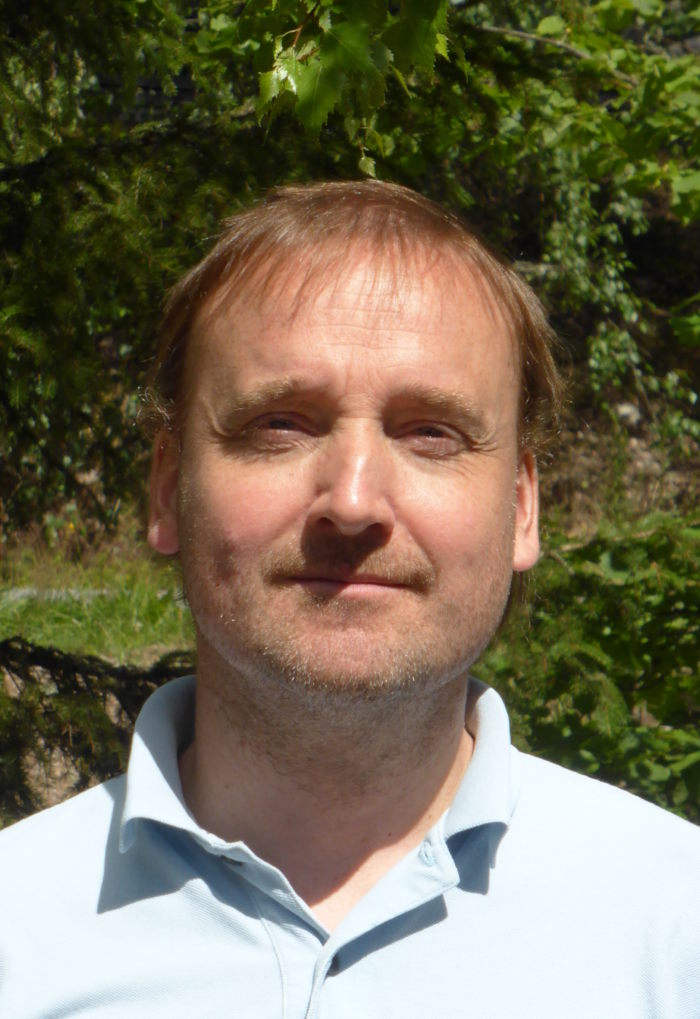}}]
{Ulrik Imberg} received the M.Sc. degree in applied physics and electrical engineering from The Institute of Technology at Linköping University in 1997. 

He is currently leading the Huawei Technologies Sweden AB Wireless Technology Planning team. 
His research interests cover 5G \& 6G telecommunication systems \& technologies, especially active antennas, antenna-near electronics and photonics.	
\end{IEEEbiography}
\begin{IEEEbiography}
 [{\includegraphics[width=1in,height=1.25in,clip,keepaspectratio]{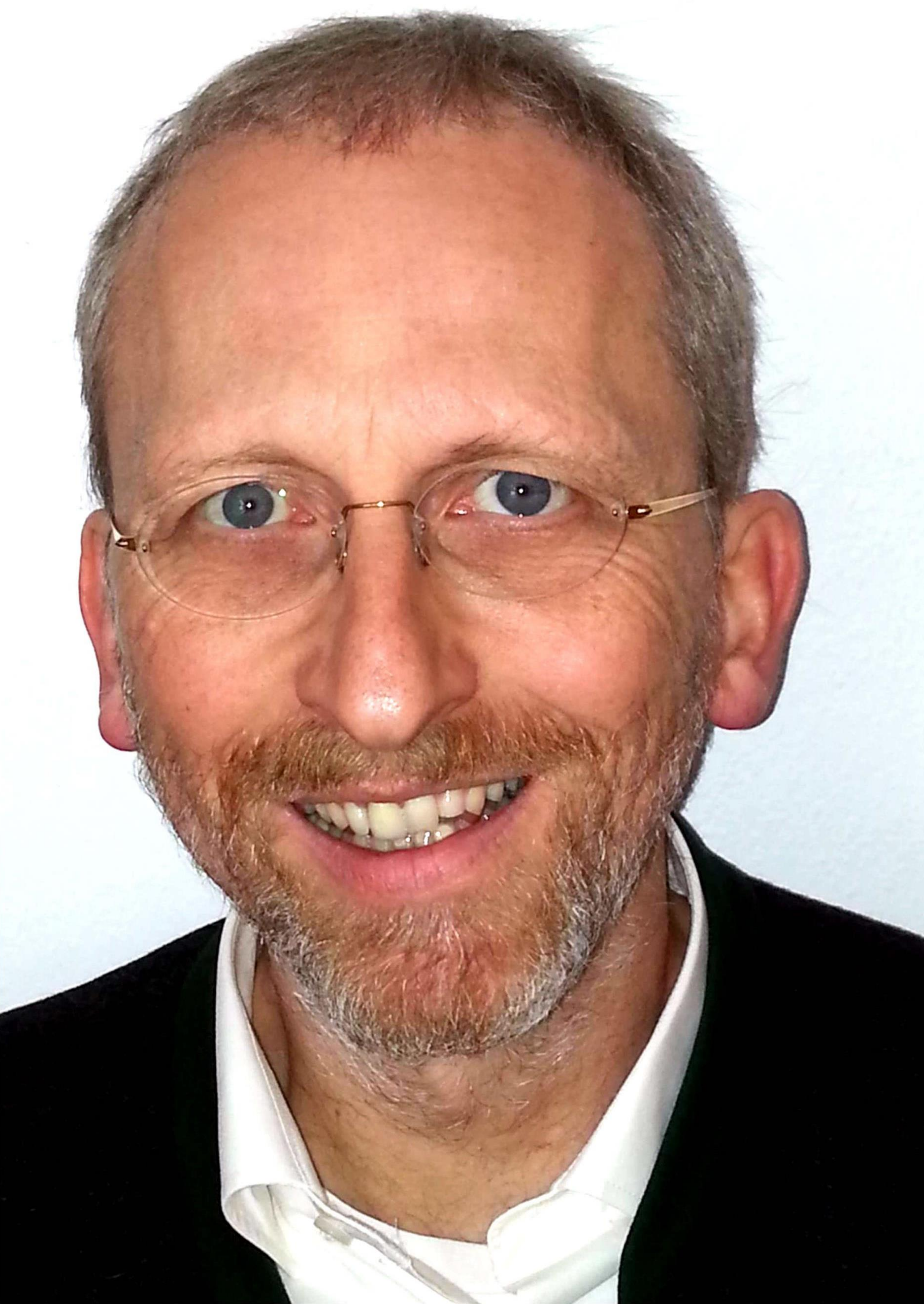}}]
{Thomas F.~Eibert} (S'93$-$M'97$-$SM'09) received the Dipl.-Ing.\,\,(FH) degree from Fachhochschule N\"urnberg, Nuremberg, Germany, the Dipl.-Ing.~degree from Ruhr-Universit\"at Bochum, Bochum, Germany, and the Dr.-Ing.~degree from Bergische Universit\"at Wuppertal, Wuppertal, Germany, in 1989, 1992, and 1997, all in electrical engineering. 

He is currently a Full Professor of high-frequency engineering at the Technical University of Munich, Munich, Germany. 
His current research interests include numerical electromagnetics, wave propagation, measurement and field transformation techniques for antennas and scattering as well as all kinds of antenna and microwave circuit technologies for sensors and communications.		
\end{IEEEbiography}
\begin{IEEEbiography}
 [{\includegraphics[width=1in,height=1.25in,clip,keepaspectratio]{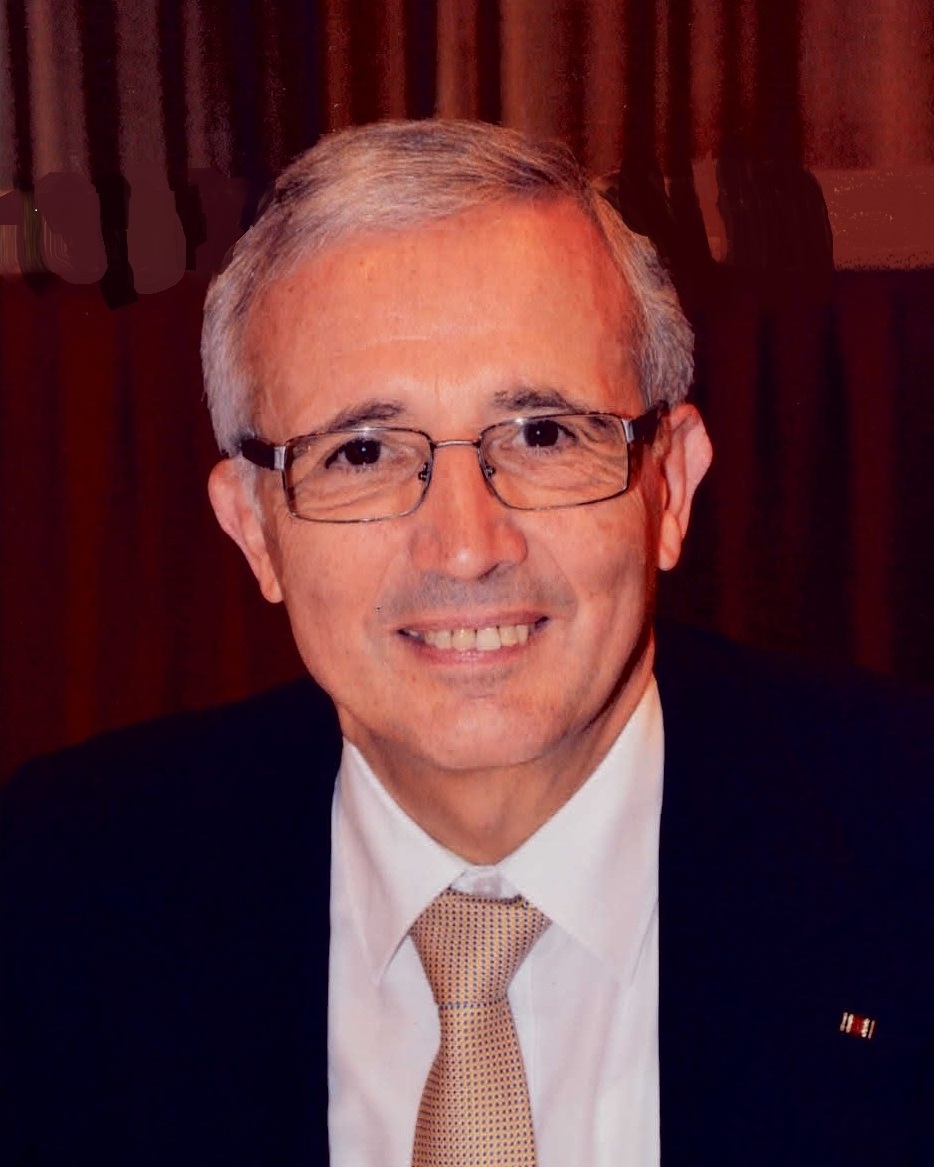}}]
{Josef A. Nossek{\normalfont~(S'72$-$M'74$-$SM'81$-$F'93$-$LF'13)}} received the Dipl.-Ing. and Dr. techn. degrees in electrical engineering from the Vienna University of Technology, Vienna, Austria, in 1974 and 1980, respectively. 

In 1974, he joined Siemens AG, Munich, Germany, where he was engaged in filter design for communication systems. 
From 1987 to 1989, he was the Head of the Radio Systems Design Department, where he was instrumental in introducing highspeed VLSI signal processing into digital microwave radio. 
From 1989 to 2016, he was a Full Professor for circuit theory and signal processing with the Technical University of Munich (TUM), Munich. 
Since 2016, he has been an Emeritus of Excellence of TUM and a Full Professor with the Federal University of Ceará, Fortaleza, Brazil. 

Dr. Nossek was the President-Elect, President, and Past President of the IEEE Circuits and Systems Society in 2001, 2002, and 2003, respectively. 
He was the President of Verband der Elektrotechnik, Elektronik, und Informationstechnik (VDE) from 2007 to 2008 and the President of the Convention of National Associations of Electrical Engineers of Europe (EUREL) in 2013. 
In 2009, he became a member of the National Academy of Engineering in Germany (acatech). 
He was a recipient of the ITG Best Paper Award in 1988, the Mannesmann Mobilfunk (currently Vodafone) Innovations Award in 1998, and the Award for Excellence in Teaching from the Bavarian Ministry for Science, Research and Art in 1998. 
From the IEEE Circuits and Systems Society, he was a recipient of the Golden Jubilee Medal for Outstanding Contributions to the Society in 1999 and the Education Award in 2008. 
He was also a recipient of the Order of Merit of the Federal Republic of Germany (Bundesverdienstkreuz am Bande) in 2008, the IEEE Guillemin-Cauer Best Paper Award in 2011, the honorary doctorate (Dr. h. c.) from the Peter Pazmany Catholic University, Hungary, in 2013, and the VDE Ring of Honor in 2014.
\end{IEEEbiography}
\vfill

\end{document}